\documentclass[draftclsnofoot,twoside,onecolumn,letter,12pt]{IEEEtran}
\usepackage{cite}
\usepackage[normalem]{ulem}
\usepackage{amsmath,amssymb,amsfonts}
\usepackage{graphicx}
\usepackage{caption}
\usepackage{textcomp}
\usepackage{xcolor}
\usepackage{graphicx}
\usepackage{amssymb}
\usepackage{amsmath}
\usepackage{cite}
\usepackage{stfloats}
\usepackage{epstopdf}
\usepackage{mdwlist}
\usepackage[section]{algorithm}
\usepackage[noend]{algpseudocode}

\usepackage[all,2cell]{xy} \UseAllTwocells
\usepackage{booktabs}

\usepackage{tikz}
\usetikzlibrary{arrows.meta}
\usetikzlibrary{decorations.pathreplacing}
\usepackage{xcolor}
\usepackage{enumitem}
\usepackage{nicefrac}

\usepackage{subcaption}
\usepackage{dsfont}
\usepackage{mathrsfs}

\newtheorem{remark}{Remark}
\newtheorem{theorem}{Theorem}
\newtheorem{proposition}{Proposition}
\newtheorem{lemma}{Lemma}

\newtheorem{proof}{Proof}
\newtheorem{example}{Example}
\newtheorem{definition}{Definition}
\newtheorem{assumption}{Assumption}

\AtBeginDocument{}

\newcommand{\beq}{\begin{equation}}
	\newcommand{\eeq}{\end{equation}}
\newcommand{\beqa}{\begin{eqnarray}}
	\newcommand{\eeqa}{\end{eqnarray}}

\newcommand{\comment}[1]{}

\DeclareMathOperator*{\argmin}{arg\,min}

\usepackage{ulem}

\newif\ifbulletlist
\newif\iftext
\texttrue 

\usepackage[affil-it]{authblk} 
\usepackage{etoolbox}
\usepackage{lmodern}

\makeatletter
\patchcmd{\@maketitle}{\LARGE \@title}{\fontsize{16}{19.2}\selectfont\@title}{}{}
\makeatother

\begin{document}
	
	\title{
		\hspace{2cm}\\[-0.8cm]
		Optimal Task Offloading Policy in Edge Computing Systems with Firm Deadlines
	}



\comment{
	\author{
		\IEEEauthorblockN{Khai Doan\IEEEauthorrefmark{1}}, 
		\IEEEauthorblockN{Wesley Araujo\IEEEauthorrefmark{1}},
		\IEEEauthorblockN{ Evangelos Kranakis\IEEEauthorrefmark{2}},\\
		\IEEEauthorblockN{ Ioannis Lambadaris\IEEEauthorrefmark{1}}, 
		\IEEEauthorblockN{Yannis Viniotis\IEEEauthorrefmark{3}}\\
		
		\IEEEauthorblockA{\IEEEauthorrefmark{1}Carleton University, Department of Systems and Computer Engineering, Ottawa, ON, Canada.}\\
		
		\IEEEauthorblockA{\IEEEauthorrefmark{2}Carleton University, School of Computer Science, Ottawa, ON, Canada.}\\
		
		\IEEEauthorblockA{\IEEEauthorrefmark{3}North Carolina State University, Department of Electrical and Computer Engineering, Raleigh, NC, USA.
		}
		
		Email: \{khaidoan@sce, wesleyaraujo@cmail, kranakis@scs, ioannis@sce\}.carleton.ca, candice@ncsu.edu }
	
\author{
	Khai Doan,
	Wesley Araujo,
	Evangelos Kranakis,\\
	Ioannis Lambadaris,
	and
	Yannis Viniotis
	\thanks{
		Khai Doan, Wesley Araujo, and Ioannis Lambadaris are with
		Carleton University, Department of Systems and Computer Engineering, Ottawa, ON, Canada.
		(e-mail: \{khaidoan@sce, wesleyaraujo@cmail, ioannis@sce\}.carleton.ca). 
	}
	\thanks{
		Evangelos Kranakis is with
		Carleton University, School of Computer Science, Ottawa, ON, Canada. (e-mail: kranakis@scs.carleton.ca).
	}
	\thanks{
		Yannis Viniotis is with North Carolina State University, Department of Electrical and Computer Engineering, Raleigh, NC, USA. (e-mail: candice@ncsu.edu).
	}
}
}

    \author{
		\IEEEauthorblockN{Khai Doan\IEEEauthorrefmark{1}, Wesley Araujo\IEEEauthorrefmark{1}, Evangelos Kranakis\IEEEauthorrefmark{2}, Ioannis Lambadaris\IEEEauthorrefmark{1}, Yannis Viniotis\IEEEauthorrefmark{3}}\\
		\IEEEauthorblockA{\IEEEauthorrefmark{1}Carleton University, Department of Systems and Computer Engineering, Ottawa, ON, Canada.}\\
		\IEEEauthorblockA{\IEEEauthorrefmark{2}Carleton University, School of Computer Science, Ottawa, ON, Canada.}\\
		\IEEEauthorblockA{\IEEEauthorrefmark{3}North Carolina State University, Department of Electrical and Computer Engineering, Raleigh, NC, USA.
		}
		
		Email: \{khaidoan@sce, wesleyaraujo@cmail, kranakis@scs, ioannis@sce\}.carleton.ca, candice@ncsu.edu }
		
	
	\maketitle
	
	
	\begin{abstract}
		The recent drastic increase in mobile data traffic has pushed the mobile edge computing systems to the limit of their capacity. A promising solution is the task migration to remote servers. Key factors to be considered in the design of offloading schemes must include the number of tasks waiting in the system as well as their corresponding deadlines. An appropriate system cost which is used as an objective function to be minimized comprises two parts. First, an offloading cost which can be interpreted as the cost of using computational resources at the external server. Second, a penalty cost due to potential task expiration. In order to minimize the expected (time average) cost over a time horizon, we formulate a Dynamic Programming (DP) equation and analyze it to describe properties of a candidate optimal offloading policy. The DP equation suffers from the well-known ``Curse of Dimensionality'' that makes computations intractable, especially when the state space is infinite. In order to reduce the computational burden, we identify three important properties of the optimal policy. Based on these properties, we show that it suffices to evaluate the DP equation on a finite subset of the state space only. We then show that the optimal task offloading decision associated with a state can be inferred from the decision taken at its ``adjacent'' states, further reducing the computational load. Finally, we provide numerical results to evaluate the influence of different parameters on the system performance as well as verify the theoretical results.
	\end{abstract}

	%
	%
	

\section{Introduction}
	
	\ifbulletlist
	{\color{red}
	    \begin{enumerate}
	        \item Explain MEC and MCC systems and their difference.
	    \end{enumerate}
	}
	\fi
	
	\iftext {
	Mobile-Edge Computing (MEC)
	and Mobile Cloud Computing (MCC) are important paradigms in addressing the limited computational capability of mobile devices. In MEC, a remote server is placed
	physically near the device for computational tasks to be
	offloaded to the MEC-server for remote computation.
	MEC is similar to MCC,
	however the server in the latter case, is not necessarily physically close to the device.
	MEC is more appropriate if the network latency or network congestion
	is a problem. In general, there are two scenarios when MEC is most appropriate:
	when the application has real-time constraints or when the user/wireless device has limited
	resources such as memory, storage, CPU, etc.
	Surveys on MEC and MCC can be found in \cite{mao2017survey} and in \cite{dinh2013survey}, respectively.
	}
	\fi
	
	\subsection{Motivation and Related Works}
	\ifbulletlist
	{\color{red}
	\begin{enumerate}
	    \item Discussion on advantages of computational offloading and some related works
	\end{enumerate}
	}
	\fi
	\iftext {
	Exploiting the advantage of offloading systems, computational services requested by users can be either processed at local servers (e.g., MEC) or offloaded onto remote servers (e.g., MCC) that have more computational capability. This feature not only improves users' quality of experience by reducing the processing, latency and power consumption, but also allows different types of applications and services to be deployed on devices with low computational capability. Due to significant advances in practical applications, analyzing MEC and MCC systems has been attracting a lot of attention in the research literature. For example, \cite{fragkos2020artificial} proposed an optimal task offloading algorithm by maximizing an appropriate utility function. Such a function increases with users' satisfaction characterized by tasks processed remotely, and decreases with respect to the total amount of computation and energy consumption overhead.
	In \cite{teymoori2020efficient}, the authors study the computational complexity of task offloading policies in a
	MCC context where tasks have hard deadlines.
	In \cite{geng2018energy}, they study the problem of computational offloading in a MCC context with
	the goal of minimizing energy consumption of the user device by taking into account its multi-core architecture.
	}
	\fi
	
	\ifbulletlist
	{\color{red}
	\begin{enumerate}
	    \item Uncertainty in the remote server availability is not well addressed in research literature, and some related works are mentioned.
	\end{enumerate}
	}
	\fi
	
	\iftext {
	In practical systems, the task offloading feature may suffer from uncertainty due to certain factors such as network congestion and limited computational resources of remote servers. Such factors may impose randomness in the availability of task migration. Systems that have firm task deadlines require more attention when task migration is not always available. These issues have not been adequately addressed in existing works which typically assume that task offloading can always be controlled.
	In \cite{zhang2014collaborative}, the authors study the problem of offloading tasks to a cloud computing
	infrastructure aiming at minimizing the energy consumption of the offloading device and meeting
		the task deadlines.
	In \cite{huang2021deadline}, the authors consider computational offloading in a real-time MEC setting in which
	tasks have hard deadlines.
	In \cite{du2017computation}, the authors study the problem of computational offloading
	in the context of a mixed MEC and MCC system
	with the goal of minimizing the weighted cost of task processing delay and energy consumption
	of the devices.
	}
	\fi
	
	\ifbulletlist
	{\color{red}
	\begin{enumerate}
	    \item There do not exist many papers on MEC or MCC that study the properties of an optimal policy.
	\end{enumerate}
	}
	\fi
	
	\iftext {
	Furthermore, despite several offloading algorithms proposing improvements for the  system performance under different contexts, to the best of our knowledge, there is a lack of studies for the the characterization of the optimal policy for task offloading to enhance the system performance.
	In \cite{van2017optimization, van2018deep}, the authors consider the problem of a user offloading tasks to multiple
	moving edge devices so as to maximize the utility of the task execution
	and minimize the energy consumption.
	In the first paper, a Deep Reinforcement Learning based method is used to solve the problem,
	while in the second paper, Q-learning is used.
	}
	\fi
	
	\subsection{Model Novelty and Main Contributions}
	\ifbulletlist
	{\color{red}
	\begin{enumerate}
	    \item Brief description of system model and optimal policy via DP equation along with contributions.
	\end{enumerate}
	}
	\fi
	\iftext {
	
	What differentiates our model from previous studies is the combination of: (a) the randomness in the connection between both the remote server and local device, (b) tasks that have firm deadlines, and, (c) the formulation of the problem in the context of  an optimal stochastic control (DP) framework.
	As is well-known in DP formulations,  computing the solution of the DP equation may become computationally prohibitive (also  known as the ``dimensionality curse'' of DP). With this motivation, our contributions can be summarized as follows.
	\begin{itemize}
		\item Our main contribution is the mathematical characterization of the structure of the optimal policy. The findings are presented in Theorems  \ref{Theo:adjacent_L*} and \ref{Theo:L*_thesmallest}. The properties we prove in these theorems help reduce the computational burden associated with the DP equation.
		\item Our second contribution is a further reduction in the computational effort for determining the minimum expected (time average) cost. In Proposition \ref{Pro:SemiRedState} we derive a mathematical property of this cost. This will enable us to compute it by applying the computationally heavy DP equation to \textit{only a finite number} of states (the so-called \textit{lean states}); the cost for an arbitrary state can then be calculated via a much simpler algebraic computation.
	\end{itemize}
	}
	\fi
	
	\ifbulletlist
	{\color{red}
	\begin{enumerate}
	    \item Organization and Sections of this paper
	\end{enumerate}
	}
	\fi
	
	\iftext {
	The rest of this article is organized as follows. In the next section, we describe the system model under consideration. In Section \ref{Sec:DP}, we formulate the DP equation to minimize the expected (time average) cost. In Section \ref{Sec:RS}, we introduce the notions of \textit{reduced} and \textit{lean state} space and show how they are used to reduce the computational load for the calculation of the optimal cost of our model. Next, we study properties of the optimal policy and the minimum expected time-average (ETA) cost in Section \ref{Sec:OptimalPolicyProp}. The optimal policy is explicitly presented in Section \ref{Sec:OptimalPolicy}. Subsequently, we illustrate numerical results in Section \ref{Sec:NumericalResults}. We conclude our work with Section \ref{Sec:Conclusion}. Finally, we provide proofs for all the theorems, lemmas, and propositions in Appendices presented in Sections \ref{Sec:AppendixTheorems}, \ref{Sec:AppendixLemmas}, and \ref{Sec:AppendixPropositions}, respectively.
	}
	\fi

	\section{System Model} \label{Sec:SystemModel}

	\begin{figure}
		\center
		\captionsetup{justification=centering}
		\includegraphics[scale=0.3]{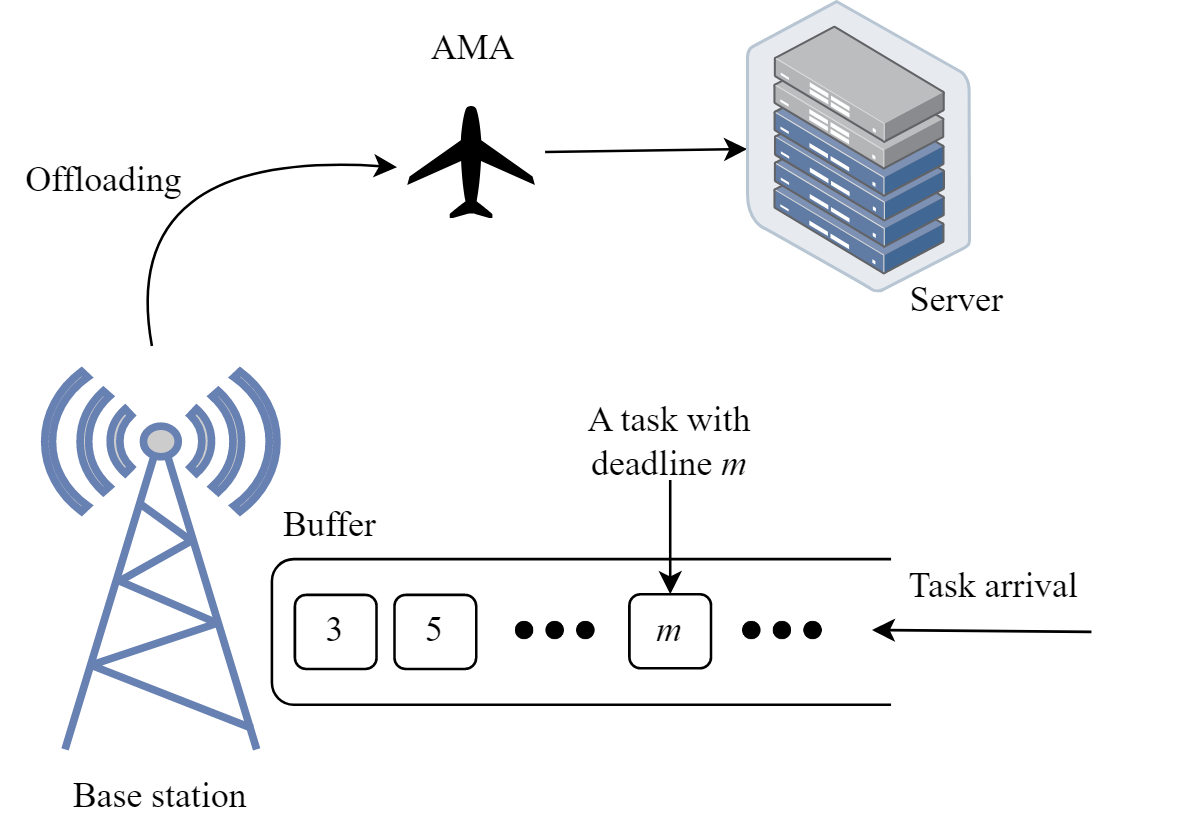}
		\caption{System model.}
		\label{fig:SM}
	\end{figure}

	The system we consider is depicted in Fig.~\ref{fig:SM}; it consists of a base station (BS) operating in discrete-time over $T$ time slots. Tasks sent by users in the area will be buffered at the BS. We assume that the maximum deadline for any task is $N$, a fixed positive integer. The deadline represents the number of time slots (including the current one) within which the tasks need to be processed, otherwise they will expire and result in a penalty $C_p$ per expired task. 
	
	We define the system state vector at a given time slot $t$ by
	\begin{align}\label{state_vec}
		\mathbf{s}_t = \left(n_1^{(t)}, n_2^{(t)}, \ldots, n_N^{(t)}\right), t \in \left\{0, 1, \ldots, T-1 \right\}
	\end{align}
	where  $n_i^{(t)}\in \mathds{Z}_+$ represents the number of tasks having deadline $i=1,2,\ldots,N$, buffered at the BS at time slot $t$. The state space for this system is $\mathds{Z}_+^N$, which is an $N$-dimensional infinite state space.
	In this system,  a number of different events takes place and triggers the system to transit to a  different state in the next time slot. Those events can be listed in order as follows.
	\begin{enumerate}
    
	\item \textbf{Connection to a remote server:} A task offloading service is provided by an external server that is outside the communication distance of the BS. The server is accessible by means of an \textit{autonomous mobile agent} (AMA). The AMA randomly arrives in the area and behaves as an intermediate node to connect the BS and the server. With the AMA, the BS can offload tasks to the server at cost $C_o > 0$ per task. Let $p_u$ be the probability that the AMA arrives in each time slot. The remote server is assumed to be capable of processing all tasks instantly without delay. In addition, we also assume that the task transferring from the BS to the AMA, and then, to the remote server incurs no delay.

	\item \textbf{The deadline shifting}: The deadlines of all tasks are reduced by one when transitioning from a time slot to the next. The deadlines are strict: a penalty of $C_p$ is incurred for every task whose deadline has dropped to 0. In this model, we assume that $C_p > C_o$.

	\item \textbf{The arrival of a new task}: During our system operation, there will be new tasks sent by users; we assume at most one new task can arrive per time slot. We use $p_i$ to denote the probability that a new task will have deadline $i$; $p_0$ denotes the probability of no task arrival. Thus, 	the task arrival probability vector is given by
	$\mathbf{p} = \left(p_0, p_1, p_2, \ldots, p_N\right).$
    We will assume that task arrivals in different time slots are independent.

	\item \textbf{The local processing service}: We assume that the BS has limited computational capability. Therefore, the local processing service provided by the BS is assumed to be available at random with probability $\mu$ in each time slot. When it is available, at most one task can be processed per slot.  
	
	\end{enumerate}

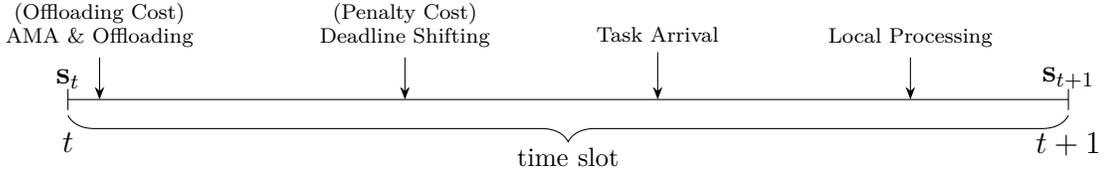
\begin{figure}
		\centering
		\begin{tikzpicture}[>=Stealth,scale=1.4]
			\draw (0,1) -- (9.5,1);
			\draw (0,0.9) -- (0,1.1);
			\draw (9.5,0.9) -- (9.5,1.1);
			\node[above] at (0,1) {$\mathbf{s}_t$};
			\node[above] at (9.5,1) {$\mathbf{s}_{t+1}$};
			\draw[decorate,decoration={brace,amplitude=10pt,mirror,raise=6pt}]
			(0,1) -- node[black,below=0.5cm] {\small time slot} (9.5,1);
			\node[below=0.3cm] at (0,1) {$t$};
			\node[below=0.3cm] at (9.5,1) {$t+1$};
			\draw[<-] (0.3,1) -- (0.3,1.4) node[above,text width=4cm,align=center,font=\scriptsize\linespread{1}\selectfont] {(Offloading Cost)\\ AMA \& Offloading};
			\draw[<-] (3.2,1) -- (3.2,1.4) node[above,text width=4cm,align=center,font=\scriptsize\linespread{1}\selectfont] {(Penalty Cost)\\ Deadline Shifting};
			\draw[<-] (5.6,1) -- (5.6,1.45) node[above] {\scriptsize Task Arrival};
			\draw[<-] (8,1) -- (8,1.4) node[above] {\scriptsize Local Processing};
		\end{tikzpicture}
		\caption{Events taking place in a time slot.} \label{fig:order-events}
	\end{figure}

	The sequence of the above events for our model is illustrated in Fig. \ref{fig:order-events}. The randomness in  local processing and in the appearance of the AMA accounts for the uncertainty in the task processing.  
	
    \ifbulletlist
    {\color{red}
    \begin{enumerate}
        \item Describing the offloading cost, penalty cost, and the number of tasks offloaded $L$.
    \end{enumerate}
    }
    \fi
    The system cost is described by the offloading  and task expiration penalty costs. To be more specific, assume that state $\mathbf{s}_t$ (as defined in Eq. (\ref{state_vec})) is encountered at time slot $t$. By $L_t$ we denote the number of tasks that are offloaded when the system is in state $\mathbf{s}_t$.   $L_t$ is also called the \textit{offloading decision}; determining its optimal value (to be defined shortly) is the subject of this work. $L_t$ can take values in the set 
	\begin{align}\label{L_definition}
		\mathbb{L}\left(\mathbf{s}_t\right) = \left\{0, 1, 2, \ldots, \sum_{i=1}^N n_i^{(t)} \right\}.
	\end{align}

    \ifbulletlist
    {\color{red}
    \begin{enumerate}
        \item Introduce the most-imminent strategy: Offloading and locally processing tasks with the smallest deadlines first.
    \end{enumerate}
    }
    \fi
    
    \iftext{
    In our model, all tasks share the same offloading cost $C_o$, and result in the same penalty $C_p$ when expire. Trivially, it is optimal to locally process and offload tasks with the most imminent deadlines first. In this work, when the AMA is available and $L \ge 1$ tasks are defined to be offloaded, we offload $L$ tasks with the smallest deadlines. Whenever the local processing is available, we process the task with the smallest deadline. The proof for this optimality property is presented in Lemma \ref{Lem:MI_optimality}.
    }
    \fi
    
    \ifbulletlist
    {\color{red}
    \begin{enumerate}
        \item Defining instantaneous cost.
    \end{enumerate}
    }
    \fi
	Based on the strategy introduced above, given state $\mathbf{s}_t$ and the offloading decision $L_t$, we define $\mathcal{C}\left(\mathbf{s}_t, L_t \right)$,  the \textit{expected instantaneous cost} incurred at time slot $t$  by:
    \begin{align} \label{eq:instant_cost}
        \mathcal{C}\left(\mathbf{s}_t, L_t \right) = p_u\mathcal{C}^{\text{A}}\left(\mathbf{s}_t, L_t \right) + \left(1 - p_u\right)\mathcal{C}^{\overline{\text{A}}}\left(\mathbf{s}_t\right)
    \end{align}
    in which
    \begin{align}\label{eq:instant_cost2}
        \mathcal{C}^{\text{A}}\left(\mathbf{s}_t, L_t \right) = C_oL_t + C_p\max\left(n_1^{(t)} - L_t, 0 \right)
    \end{align}
    is the instantaneous cost when the AMA arrives and $L_t$ tasks are offloaded. Similarly,
    \begin{align}\label{eq:instant_cost3}
        \mathcal{C}^{\overline{\text{A}}} = C_pn_1^{(t)}
    \end{align}
    is the cost when AMA is not available, thus, does not depend on the offloading decision $L_t$.
	
    For a given time horizon $T$, an offloading {\em policy $\pi$} is a rule that determines the offloading parameter $L_t$ for every state $\mathbf{s}_t$ at $t$, i.e., $L_t=\pi(\mathbf{s}_t)$.

	\section{Dynamic Programming Formulation } \label{Sec:DP}
	\ifbulletlist
    {\color{red}
    \begin{enumerate}
        \item Defining the expected time-average cost.
    \end{enumerate}
    }
    \fi

    \iftext{
    Consider the expected time-average cost  given by
	\begin{equation}\label{eq:sum_cost}
         \frac{1}{T}
		\sum_{t=0}^{T-1} \mathds{E}[\mathcal{C}\left(\mathbf{s}_t, L_t\right)].
	\end{equation}
    
  For the rest of this work, we will refer to the cost defined above as the \textit{average cost}. We wish to determine the optimal offloading policy that minimizes the above average cost for a given time horizon.
    }
    \fi
    
	\ifbulletlist
	{\color{red}
	\begin{enumerate}
	    \item New notation for state transition $\mathbf{s}'_{Lk}$ and $\mathbf{s}''_{Lk}$.
	\end{enumerate}
	}
	\fi
	
	\iftext{
	Consider a state $\mathbf{s}$, in which the decision is to offload $L$ tasks. Suppose a new task arrives with a deadline $k$. If  local processing is available, the system will transit to a new state  $\mathbf{s}'_{Lk}$; if not, the new state will be $\mathbf{s}''_{Lk}$. 
	These state definitions facilitate the analysis of the Dynamic Programming equation that will follow. They are formally defined in Eqs. (\ref{next_state}) in this section. 	We need to introduce some additional, detailed notation to that effect. 
	}
	\fi
	
	\ifbulletlist
	{\color{red}
	\begin{enumerate}
	    \item Example of system evolution for intuition of notation $\mathbf{s}'_{Lk}$ as well as the transition probability function.
	\end{enumerate}
	}
	\fi
	
	\iftext{
Before doing this, we  provide an example, let $\mathbf{s} = (0,1,2,0,1)$, the next event according to Fig. \ref{fig:order-events} is offloading, the AMA is available and $L=2$ tasks with the most imminent deadlines are to be offloaded, therefore the intermediate state is $(0,0,1,0,1)$. The next event is deadline shifting to account for the new deadlines of the tasks in the next time slot, resulting in $(0,1,0,1,0)$. Following Fig. \ref{fig:order-events}, we assume a new task with deadline 3 arrives, resulting in the intermediate state $(0,1,1,1,0)$. The next event is local processing, which processes the most imminent task, resulting in system state  $\mathbf{s}'_{23} = (0,0,1,1,0)$.
	}
	\fi
	
	\ifbulletlist
    {\color{red}
    \begin{enumerate}
        \item Defining vector $\mathbf{a}_k$ representing the event of task arrival with deadline $k$.
    \end{enumerate}
    }
    \fi
    
    \iftext{
    Vector $\mathbf{o}\left(\mathbf{s}, L\right) = \left(o_1, \ldots, o_N\right)$ represents the task offloading. Let $d'$ be the deadline of the most imminent task after $L$ tasks have been offloaded from $\mathbf{s}$. If $L < n_1$, $d' = 1$. Otherwise, $d'$ satisfies the inequalities 
    \begin{align}\label{ineq:d'_cond2}
        L \le \sum_{j=1}^{d'} n_j \text{ and } L > \sum_{j=1}^{d'-1} n_j.
    \end{align}
    Then, $\mathbf{o}\left(\mathbf{s}, L\right)$ is defined as follows. 
    \begin{itemize}
        \item If $d' = 1$:
        \begin{align}
        o_i =
            \begin{cases}
                L, & \text{ if } i = 1,\\
                0, & \text{ if } i \ge 2.
            \end{cases}
        \end{align}
    
        \item If $d'$ satisfies inequalities (\ref{ineq:d'_cond2}):
        \begin{align}
        o_i =
            \begin{cases}
                n_i, & \text{ if } i \le d'-1,\\
                L -  \sum_{j=1}^{d'-1} n_j, & \text{ if } i = d',\\
                0, & \text{ if } i \ge d'+1.
            \end{cases}
        \end{align}
    \end{itemize}
    Let $\mathbf{a}_k = \left(a_1, \ldots, a_N \right)$ represent the task arrival vector defined as:
    \begin{align}
        a_i =
        \begin{cases}
            1 &, \text{ if } i = k,\\
            0 &, \text{ if } i \ne k.
        \end{cases}
    \end{align}
    Then, the intermediate state of the system after $L$ most imminent tasks have been offloaded from $\mathbf{s}$, the deadline shifting has been performed, and the task arrival event has been realized, can be defined as follows:
    \begin{align}
        \mathbf{s}^{inte} = f_{ds}\left(\mathbf{s} - \mathbf{o}\left(\mathbf{s}, L\right)\right) + \mathbf{a}_k
    \end{align}
    where the function $f_{ds}\left(\cdot\right)$ performs deadline shifting on the given vector. If $\mathbf{s} = \left(n_1, \ldots, n_N\right)$, $f_{ds}\left(\mathbf{s}\right)$ is an $N$-dimensional vector $\left(n_2, \ldots, n_N, 0\right)$ in which the component $n_1$ is removed, representing task expiration.

    Let $\mathbf{l} = \left(l_1, \ldots, l_N\right)$ represent the local processing vector. If $\mathbf{s}^{inte} \ne (0, \ldots, 0)$, we assume $d$ is the deadline of the most imminent task in $\mathbf{s}^{inte}$, then elements of $\mathbf{l}$ is given by:
    \begin{align}
        l_i =
        \begin{cases}
            1, & \text{ if } i = d,\\
            0, & \text{ if } i \ne d.
        \end{cases}
    \end{align}
    Otherwise, $\mathbf{l}=(0, \ldots, 0)$.
	}
	\fi
	
	\ifbulletlist
    {\color{red}
    \begin{enumerate}
        \item Defining possible transition states in the next time slot and transition probability function.
    \end{enumerate}
    }
    \fi
    
    \iftext{
    Then the system state transition is defined as:
	\begin{align} \label{next_state}
		\mathbf{s}'_{Lk} = \mathbf{s}^{inte} - \mathbf{l}, \text{ and } \mathbf{s}''_{Lk} = \mathbf{s}^{inte}.
	\end{align}
	}
	\fi
	
	\ifbulletlist
	{\color{red}
	\begin{enumerate}
	    \item Expressing the DP equation.
	\end{enumerate}
	}
	\fi
	
	\iftext{
	To this end, let $J_T\left(\mathbf{s}\right)$ denotes the minimum average cost over $T$ time slots for a given initial state $\mathbf{s}$. We have the following Dynamic Programming Equation.
	\begin{align}
		\label{DP_Eq}
		\begin{split}
			&J_{T}\left( \mathbf{s} \right) = ~ \underset{L \in \mathbb{L}\left(\mathbf{s}\right)}{\min}~ \left\{\mathcal{C}\left(\mathbf{s}, L\right) + G_{T-1}\left(\mathbf{s}, L\right)\right\},
		\end{split}
	\end{align}
	with
	\begin{align} \label{G_func}
		G_T\left(\mathbf{s}, L\right) =
		\begin{cases}
			p_u G_T^{{\text{A}}}\left(\mathbf{s}, L\right) + \left(1-p_u\right)G_T^{\overline{\text{A}}}\left(\mathbf{s}\right), & T \ge 1,\\
			0, &  T = 0.
		\end{cases}
	\end{align}
	where 
	
	\begin{align}\label{G_AMA}
		G_T^{{\text{A}}}\left(\mathbf{s}, L\right) =   \mu \sum_{k=0}^N p_k J_T\left(\mathbf{s}'_{Lk} \right)   + \left(1-\mu\right)\sum_{k=0}^N p_k J_T\left(\mathbf{s}''_{Lk}\right)
	\end{align}
	is the average future cost given that the AMA arrives at the current time slot and $L$ tasks are offloaded, and
	\begin{align}\label{G_NAMA}
		G_T^{\overline{\text{A}}}\left(\mathbf{s}\right) =   \mu \sum_{k=0}^N p_k J_T\left(\mathbf{s}'_{0k} \right)   + \left(1-\mu\right)\sum_{k=0}^N p_k J_T\left(\mathbf{s}'_{0k} \right)
	\end{align}
	is that without the AMA, and hence, does not depend on $L$.
	}
	\fi

 Eq. (\ref{DP_Eq}) is equivalent to
	\begin{align}\label{gen_U_noU}
		J_{T+1}\left(\mathbf{s}\right) = p_uJ_{T+1}^{\text{A}}\left(\mathbf{s}\right) + \left(1-p_u\right)J_{T+1}^{\overline{\text{A}}}\left(\mathbf{s}\right)
	\end{align}
	in which
	\begin{align}\label{eq:JAMA}
		J_{T+1}^{\text{A}}\left(\mathbf{s}\right) = \underset{L \in \mathbb{L}\left(\mathbf{s}\right)}{\min}\left\{\mathcal{C}^{\text{A}}\left(\mathbf{s}, L\right) + G_T^{\text{A}}\left(\mathbf{s}, L\right) \right\}
	\end{align}
	and
	\begin{align}\label{eq:JNoAMA}
		J_{T+1}^{\overline{\text{A}}}\left(\mathbf{s}\right) = \mathcal{C}^{\overline{\text{A}}}\left(\mathbf{s}\right) + G_T^{\overline{\text{A}}}\left(\mathbf{s}\right).
	\end{align}
	
	To facilitate our analysis in the subsequent sections, we define the minimum average cost attained by offloading exactly $L$ most imminent tasks from an initial state $\mathbf{s}$ given the AMA's presence by
	\begin{align}\label{eq:JsL_def}
		J_T^{\text{A}}\left(\mathbf{s}, L\right) = \mathcal{C}^{\text{A}}\left(\mathbf{s}, L\right) + G_T^{\text{A}}\left(\mathbf{s}, L\right).
	\end{align}
	We denote $\bar{\mathbf{s}}_{dL}$ the state obtained by offloading, from $\mathbf{s}$, $L$ most imminent tasks having deadline greater than or equal to $d$. The following example is to provide more intuition about this notation.
\begin{example}
Given state $\mathbf{s} = \left(0, 5, 6, 7, 8\right)$, the state $\bar{\mathbf{s}}_{37}$ is obtained by offloading 7 most imminent tasks starting from deadline 3, thus, $\bar{\mathbf{s}}_{37} = \left(0, 5, 0, 6, 8 \right)$. Now, if we offload 7 most imminent tasks starting from deadline 5, the resulted state would be $\bar{\mathbf{s}}_{57} = \left(0, 5, 6, 7, 1 \right)$.
\end{example}

We have mentioned in the previous section that whenever offloading is possible, and $L$ tasks need to be offloaded from a given state $\mathbf{s}$, it is optimal to offload $L$ most imminent tasks. This offloading behaviour results in state $\bar{\mathbf{s}}_{1L}$ which is always associated with $d=1$. However, the properties we studied in Subsection \ref{SubSec:Convex} are associated with a general context where tasks can be offloaded starting from an arbitrary deadline $d$. Therefore, for consistency, we introduce a general notation  $\bar{\mathbf{s}}_{dL}$ where $d$ is used in the subscript instead of 1. 
	
	For the rest of this article, the terms \textit{offloading states} and \textit{non-offloading states} will be used. Therefore, they are specified in the following definition.
	\begin{definition}\label{Def:Of_NOf}
		\textit{For a given state $\mathbf{s}$ and a time horizon $T$, $\mathbf{s}$ is called an offloading state if the associated optimal offloading decision is a positive integer. State $\mathbf{s}$ is called a non-offloading state if the associated optimal offloading decision is 0.}
	\end{definition}
	
	\comment{
	To this point, we would like to clarify that for every given state vector $\mathbf{s}$ and time horizon $T$, tasks having deadlines greater than $T$ will not expired within the considered time horizon. Therefore, it is trivially that these tasks will not be offloaded by the optimal policy, and hence, will not contribute to the overall cost. From this, the following assumption is made.
	\begin{assumption}\label{assumption_NT}
		\textit{For every given state $\mathbf{s}$ and time horizon $T$, the dimension of $\mathbf{s}$ is assumed not to exceed $T$.}
	\end{assumption}
	To be clear, in (\ref{DP_Eq}), the dimension of $\mathbf{s}$ is assumed not to exceed $T+1$. Meanwhile, in (\ref{G_AMA}) and (\ref{G_NAMA}), the dimensions of $\mathbf{s}'_{Lk}$, $\mathbf{s}''_{Lk}$, $\mathbf{s}'_{0k}$, and $\mathbf{s}''_{0k}$ are assumed not to exceed $T$. 
	}

	\section{Computational Load Reduction} \label{Sec:RS}
	
	\subsection{Reduced State Space}
	\ifbulletlist
	{\color{red}
		\begin{enumerate}
			\item Brief idea of reduced states.
		\end{enumerate}
	}
	\fi
	\iftext{
		For a given $N$-dimensional state vector $\mathbf{s}$, since there are at most $N-1$ tasks that can be processed in $N$ time slots. This is because the deadline shifting happens at the beginning of every time slot as presented in Fig~\ref{fig:order-events}. Therefore, tasks having deadline 1 cannot be processed. This means that there might be certain tasks that are {\em guaranteed to expire} if not offloaded within the next $N$ time slots; we will call such tasks \textit{excessive tasks}. We define the \textit{reduced states} as the ones having no excessive tasks.
	}
	\fi
	
	\ifbulletlist
	{\color{red}
		\begin{enumerate}
			\item Mathematical description of reduced state.
		\end{enumerate}
	}
	\fi
	
	\iftext{
		In the remainder of this sub-section we will provide a characterization and properties of reduced states. For such a state $\mathbf{s} = \left(n_1, \ldots, n_N\right)$ and from Fig.~\ref{fig:order-events}, all tasks having deadline 1 will expire if not offloaded, hence, we must have $n_1=0$. Next, at most one task can be processed in the next slot, therefore, we must have $n_1 + n_2 \le 1$. Subsequently, at most two tasks can be processed in the next two time slots, leading to $n_1 + n_2 + n_3 \le 2$, otherwise at least one task will expire  after two slots, and so on. Finally, at most $N-1$ tasks can be processed in $N$ time slots including the current one, so we must have $\sum_{i=1}^{N} n_i \le N-1$. The definition for a reduced states follows.
		\begin{definition}\label{Def:ReducedStates}
			\textit{
				A state $\mathbf{s}_r = \left(n^r_1, \ldots, n^r_N\right)$ is a reduced state if and only if the following inequalities hold: 
				\begin{align}\label{ineq:reduced_cond}
					\sum_{i=1}^m n_i^r \le m - 1, m=1, 2, \ldots, N.
				\end{align}
			} 
			
		\end{definition}
	}
	\fi
	
	\ifbulletlist
	{\color{red}
		\begin{enumerate}
			\item Number of possible reduced states follows Catalan number \sout{and transition to lean states}
		\end{enumerate}
	}
	\fi
	
	\iftext{
		As elements of a reduced state vector is bounded, the number of reduced vectors is finite. Moreover, the next lemma states that the number of reduced state vectors is equal to the Catalan number \cite{stanley2015catalan}.
		\begin{lemma}\label{lem_Catalan}
			The number of reduced states having dimension $N$ is finite and equals to the Catalan number $C_N = \binom{2N}{N}/(N+1)$.
		\end{lemma}
		\textit{Proof}: Please see Appendix \ref{proof_lem_Catalan}.
		
		Abusing the notation slightly, for the sake of simplicity,  we can associate a  corresponding reduced state $\mathbf{s}_r = \left(n^r_1, \ldots, n^r_N\right)$ for any given state $\mathbf{s} = \left(n_1, \ldots, n_N\right)$, in the infinite state space, using Algorithm \ref{Gen2Red_Algo}. For $\mathbf{s}=\left(n_1, \ldots, n_N\right)$, it can be seen from Eq. (\ref{DP_Eq}) and Eqs. (\ref{eq:instant_cost})-(\ref{eq:instant_cost3}) that $n_i, i \ge 2$ do not contribute to the cost $J_1\left(\mathbf{s}\right)$. Similarly, $n_i, i\ge 3$ do not contribute to the cost $J_2\left(\mathbf{s}\right)$. Observe, therefore that in general, tasks having deadline greater than the considered time horizon will not contribute to the cost in Eq. (\ref{DP_Eq}). Therefore, these tasks will not be considered, which is reflected by line 4 of this algorithm.
		
		\begin{algorithm} 
		\caption{Derivation of Reduced States} \label{Gen2Red_Algo}
		\begin{algorithmic}[1] 
			\State \textbf{Input}: $\mathbf{s} = \left(n_1, n_2, \ldots, n_N\right)$, $N$, $T$.
			\State \textbf{Output}: $\mathbf{s}_r$, $L_g$. 
			\State Initialize: $t_{\text{idle}} \leftarrow 0$, $L_g \leftarrow 0$, $\tilde{s} \leftarrow \left(\tilde{n}_1, \ldots, \tilde{n}_N\right)$ where $\tilde{n}_i \leftarrow n_i, i \le T$, and $\tilde{n}_i \leftarrow 0, i > T$.
            \State $n_i \leftarrow 0$, for $i > \min\left(N, T\right).$
			\For{$i=1,2,\ldots,N$}
			\For{$j = 1, 2, \ldots, \min\left(N, T\right)$}
			\State $a \leftarrow \sum_{d=1}^j \tilde{n}_d - \max\left(n_i - i + t_{\text{idle}} + 1, 0\right)$.
			\If{$a \ge 0$}
			\If{$j \ge 2$}
				\State $\tilde{n}_d \leftarrow 0, d = 1, \ldots, j-1$.
			\EndIf
			\State $\tilde{n}_j \leftarrow \tilde{n}_j - a$.
			\EndIf
			
			\EndFor
			
			\State $L_g \leftarrow L_g + \max\left(n_i - i + t_{\text{idle}} + 1, 0\right)$.
			\State $t_{\text{idle}} \leftarrow \sum_{d=1}^i \tilde{n}_d$.
			\EndFor
			\State $\mathbf{s}_r \leftarrow$ Offloading $L_g$ most imminent tasks from $\mathbf{s}$.
            \State \textbf{return} $\mathbf{s}_r$, $L_g$.
		\end{algorithmic}
	\end{algorithm}
		
		Determining the reduced states is the first step in reducing the computational burden of the DP equation~(\ref{DP_Eq}). The next step is the determination of a new concept know as the lean states, which also constitute a finite set and are directly used to compute the optimal cost $J_T\left(\mathbf{s}\right)$ according to the forthcoming Proposition~\ref{Pro:SemiRedState}.
	}
	\fi
	
	
	\subsection{Lean State Space} \label{sec:semi-reduced}
	\ifbulletlist
	{\color{red}
		\begin{enumerate}
			\item Brief intuition followed by a formal description of lean states
		\end{enumerate}
	}
	\fi
	\iftext{
		The definition of lean states is given as below:
		\begin{definition}\label{Def:SemRedStates}
			\textit{For a given state $\mathbf{s} = \left(n_1, \ldots, n_N\right)$, we call $\mathbf{s}_r = \left(n_1^r, \ldots, n_N^r\right)$ a reduced state obtained from $\mathbf{s}$ following Algorithm \ref{Gen2Red_Algo}. We define the parameters $\gamma_j, j = 1, \ldots, N$ as follows:} 
			\begin{align}\label{eq:gamma_def}
				\gamma_j = 
				\begin{cases}
					0 &, \text{ if } j = 1 \text{ or } j > \min\left(N, T\right), \\
					\min\left\{n_j, j - 1 - \sum_{i=1}^{j-1} \gamma_i \right\} &, \text{ otherwise}.
				\end{cases}
			\end{align}
			\textit{Then, the lean state $\mathbf{s}_\text{m}= \left(n_1^m, \ldots, n_N^m\right)$ corresponding to $\mathbf{s}$ is given by}
			\begin{align}\label{eq:semi_element_def}
			n^{\textit{m}}_i = \max\{ \gamma_i, n^\textit{r}_i\}, ~ i=1, 2, \ldots, N.
			\end{align}
		\end{definition}
	}
	\fi
	
	\ifbulletlist
	{\color{red}
		\begin{enumerate}
			\item Important property/result of lean states.
		\end{enumerate}
	}
	\fi
	
	\iftext{

		The relation in the minimum average cost of a given state $\mathbf{s}$ and that of its corresponding lean state is presented in Proposition \ref{Pro:SemiRedState}.
		
		\begin{proposition} \label{Pro:SemiRedState}
			\textit{Given state $\mathbf{s} = \left(n_1, \ldots, n_N\right)$ and its lean state $\mathbf{s}_m = \left(n_1^m, \ldots, n_N^m\right)$. Let $P_t$ be the probability that the AMA is available for the
				first time at time slot $t$. Then, the following equality holds.}
			\begin{equation} \label{eq:gen2sem_eq}
				J_T\left(\mathbf{s}\right) = J_T\left(\mathbf{s}_m\right) + C_{g2m}
			\end{equation}
			\textit{where }
			\begin{equation*}
				C_{g2m} = C_o\sum_{i=1}^{N} \left(n_i - n_i^m \right)  \sum_{j=0}^{i-1} P_{t=j}+ C_p\sum_{i=1}^{N-1} P_{t=i} \sum_{j=1}^{i}  \left(n_j - n_j^m \right) + C_pP_{t \ge N}\sum_{j=1}^{N} \left(n_j- n_j^m\right)
			\end{equation*}
			\textit{with the probabilities $P_{t=j}$ and $P_{t \ge N}$ are computed by}
			\begin{align*}
				P_{t=j} &= p_u\left(1-p_u\right)^{j},\\
				P_{t \ge N} &=  \left(1-p_u \right)^{N}.
			\end{align*}
		\end{proposition}
		\textit{Proof}: Please see Appendix \ref{proof_Pro:SemiRedState}.
		
		As the lean state space is finite, if $J_T\left(\mathbf{s}_m\right)$ is computed for all lean states via the DP equation (\ref{DP_Eq}), $J_T\left(\mathbf{s}\right)$ can be computed for every $\mathbf{s}$ via the simple algebraic manipulation of equation (\ref{eq:gen2sem_eq}).
	}
	\fi

    \section{Properties of Optimal Policy and Cost Function} \label{Sec:OptimalPolicyProp}
	\subsection{Optimality of the Most-Imminent Offloading Method}  \label{Subsec:Optim_MI}
	The first property of an optimal offloading policy is offloading the most imminent tasks first. This property will be proved based on another property presented in the next proposition.
	\begin{proposition} \label{Pro:deadline_postponed}
		\textit{Given state $\mathbf{s} = \left(n_1, n_2, \ldots, n_N \right) \not\equiv \left(0, \ldots, 0\right)$. Assuming that $d$ is the deadline of the most imminent task of $\mathbf{s}$, we define the set $\mathbb{S}_{pp}\left(\mathbf{s}\right)$ as follows. If $\tilde{\mathbf{s}} = \left(\tilde{n}_1, \ldots, \tilde{n}_N \right) \in \mathbb{S}_{pp}\left(\mathbf{s}\right)$, state $\tilde{\mathbf{s}}$ satisfies the following conditions:
		\begin{itemize}
			\item If $N = 1$: $\tilde{\mathbf{s}} \equiv \mathbf{s}$.
			\item If $N \ge 2$: either $\tilde{\mathbf{s}} \equiv \mathbf{s}$, or $\tilde{\mathbf{s}}$ is defined by
			\begin{align}
				\tilde{n}_i =
				\begin{cases}
					n_d - 1, & \text{ if } i = d,\\
					n_j + 1, & \text{ if } i = \tilde{d} \text{ for a deadline } \tilde{d} \in \left\{d+1, \ldots, N\right\}, \\
					n_i, & \text{ otherwise.}
				\end{cases}
			\end{align}
		\end{itemize}
			Then, the following inequality holds:
			\begin{align}
				J_T\left(\mathbf{s}\right) \ge J_T\left(\tilde{\mathbf{s}}\right), \text{ for all } T, \mathbf{s}, \tilde{\mathbf{s}} \in \mathbb{S}_{pp}\left(\mathbf{s}\right).
			\end{align}
		}
	\end{proposition}
	\textit{Proof}: Please see Appendix \ref{proof_Pro:deadline_postponed}.
	
	As a result of Proposition \ref{Pro:deadline_postponed}, the first property of the optimal policy can be proved, and formally presented in Lemma \ref{Lem:MI_optimality}.
	
	\begin{lemma}\label{Lem:MI_optimality}
		\textit{When processing is possible, it is optimal to process the most imminent task. When  it is optimal to offload $L$ tasks,   the $L$  most imminent tasks  should be offloaded.}
	\end{lemma}
	\textit{Proof}: Please see Appendix \ref{proof_Lem:MI_optimality}.
	
	\subsection{Convexity of the Minimum average cost with respect to the Offloading Decision} \label{SubSec:Convex}
	For a given time horizon $T$, and an initial state $\mathbf{s} = \left(n_1, \ldots, n_N\right)$, we define the following function
	\begin{align} \label{F_def}
		F^{\overline{\text{A}}}\left(T, \mathbf{s}, d, L\right) = J_T^{\overline{\text{A}}}\left(\bar{\mathbf{s}}_{dL}\right) + LC_o,
	\end{align}
	in which we recall that $\bar{\mathbf{s}}_{dL}$ is the state obtained by offloading, from $\mathbf{s}$, $L$ most imminent tasks having deadline greater than or equal to $d$. Function $F^{\overline{\text{A}}}\left(T, \mathbf{s}, d, L\right)$ is characterized by parameters $T, \mathbf{s}$, and $d$. This function takes $L$ as variable. We define the domains of $F^{\overline{\text{A}}}\left(T, \mathbf{s}, d, L\right)$ as follows.
	\begin{align}
	\mathbb{L}_1\left(\mathbf{s}\right) &= \left\{n_1, \ldots, \sum_{i=1}^N n_i \right\}, \label{F_domain1}\\
	\mathbb{L}_d\left(\mathbf{s}\right) &= \left\{0, 1, \ldots, \sum_{i=d}^N n_d \right\}, \text{ for } d = 2, \ldots, N. \label{F_domain2}
	\end{align}
	$\mathbb{L}_d\left(\mathbf{s}\right)$ can be interpreted as the set of valid offloading decisions associated with state $\mathbf{s}$ and deadline $d$. In the definition of $\mathbb{L}_1\left(\mathbf{s}\right)$, because all tasks having deadline 1 are excessive tasks and must be offloaded, the optimal offloading decision must be conveyed in the set $\left\{n_1, \ldots, \sum_{i=1}^N n_i \right\}$. Hence, $L <  n_1$ can be removed from the domains of $F^{\overline{\text{A}}}\left(T, \mathbf{s}, d, L\right)$ when $d=1$. 
	\begin{example}
	As an example, let us consider a 3-dimensional state vector $\mathbf{s} = \left(2, 3, 4\right)$. In this example, we have
	\begin{align}
			\mathbb{L}_1\left(\mathbf{s}\right) &= \left\{2, 3, \ldots, 9 \right\},\\
			\mathbb{L}_2\left(\mathbf{s}\right) &= \left\{0, 1, \ldots, 7 \right\},\\
			\mathbb{L}_3\left(\mathbf{s}\right) &= \left\{0, 1,  \ldots, 4 \right\}.
	\end{align}
	\end{example}
	
	We have the following convexity property stated in the next lemma.
	\begin{lemma}\label{Lem:Convexity}
		\textit{For every given time horizon $T$, an initial state $\mathbf{s}$, and a deadline $d$, the function $F^{\overline{\text{A}}}\left(T, \mathbf{s}, d, L\right)$ as defined in Eq. (\ref{F_def}) is a discrete convex function with respect to $L \in \mathbb{L}_d\left(\mathbf{s}\right)$ where $\mathbb{L}_d\left(\mathbf{s}\right)$ is defined in Eqs. (\ref{F_domain1}) and (\ref{F_domain2}).}
	\end{lemma}
	\textit{Proof}: Please see Appendix \ref{proof_Lem:Convexity}.
	
	The next lemma states the relation between function $F^{\overline{\text{A}}}\left(T, \mathbf{s}, d, L\right)$ and the optimal offloading decision associated with state $\mathbf{s}$ and a time horizon $T$.
	\begin{lemma}\label{Lem:Convex2L*}
		\textit{ Assuming that the function $F^{\overline{\text{A}}}\left(T, \mathbf{s}, d, L\right)$ attains its minimum at $L^*$ for $d=1$, then, $L^*$ is the optimal offloading decision of $\mathbf{s}$ for a time horizon $T$.}
	\end{lemma}
	\textit{Proof}: Please see Appendix \ref{proof_Lem:Convex2L*}.
	
	Other important properties of the optimal offloading decisions will be presented in the next subsection based on the concept of \textit{adjacent states}.
	
	\subsection{Adjacent States}
	The goal of this subsection is introducing the concept of adjacency among states, and related properties. These properties facilitate the design of the optimal policy presented later on. The definition of adjacent states is given below.
	\begin{definition}\label{Def:adjacent_states}
		\textit{Consider a state $\mathbf{s} = \left(n_1, \ldots, n_N \right) \ne \left(0, \ldots, 0 \right)$, with $d$ as the smallest deadline such that $n_d > 0$. Then, state $\mathbf{s}_a = \left(n^a_1, \ldots, n^a_N\right)$ is an adjacent state to $\mathbf{s}$ if there exists a deadline $j \in \left\{1, \ldots, d \right\}$ such that:} 
		\begin{align}
		n_i^a =
		\begin{cases}
		n_i + 1, & \text{ if } i = j,\\
		n_i, & \text{ if } i \ne j.
		\end{cases}
		\end{align}
		
		\textit{If a state $\mathbf{s}_a$ has only one task with arbitrary deadline, it is adjacent to state $\left(0, \ldots, 0 \right)$.}
	\end{definition}

	\ifbulletlist
	{\color{red}
	\begin{enumerate}
	    \item The relation between optimal decisions of two adjacent states.
	\end{enumerate}
	}
	\fi
	
	\iftext{
	Let us consider the following two examples. 
	\begin{example}
	In the first example, we assume that a state $\mathbf{s} = \left(0, 0, 1, 4, 4 \right)$ is given in which the deadline of the most imminent task in $\mathbf{s}$ is 3. Therefore, an adjacent state $\mathbf{s}_a$ of $\mathbf{s}$ can be obtained by adding a task with deadline less than or equal to 3, e.g., $\mathbf{s}_a = \left(0, 1, 1, 4, 4\right)$. 
	
	In the second example, assuming that $\mathbf{s}_a = \left(0, 2, 1, 3, 3\right)$. A state $\mathbf{s}$ for which $\mathbf{s}_a$ is adjacent, can be obtained by offloading the most imminent task in $\mathbf{s}_a$, i.e., $\mathbf{s} = \left(0, 1, 1, 3, 3\right)$.
	\end{example}
	
	We denote by $\mathbb{S}_{adj}\left(\mathbf{s}\right)$ the set of all adjacent states of $\mathbf{s}$. For a given time horizon, the optimal offloading decision of $\mathbf{s}$ can be inferred from that of $\mathbf{s}_a \in \mathbb{S}_{adj}\left(\mathbf{s}\right)$ and vice versa, as described in Theorem \ref{Theo:adjacent_L*}.
	\begin{theorem}\label{Theo:adjacent_L*}
        \textit{Given two states $\mathbf{s}$, $\mathbf{s}_a \in \mathbb{S}_{adj}\left(\mathbf{s}\right)$, and a time horizon $T$. We call $L^*$ and $L^*_a$ the optimal offloading decision of $\mathbf{s}$ and $\mathbf{s}_a$, respectively, for the time horizon $T$. We have the following relations:}
		\begin{enumerate}
			\item \textit{If $L_a^* \ge 1, L^* = L_a^*-1$.}
			\item \textit{If $L_a^* = 0$, $L^*=0$.}
			\item \textit{If $L^* \ge 1$,  $L^*_a = L^* + 1$.}
		\end{enumerate}
	\end{theorem}
	\textit{Proof}: Please see Appendix \ref{proof_Theo:adjacent_L*}.
	}
	\subsection{Offloading and Non-Offloading Conditions}\label{Subsec:Conditions}
	We have defined the notion of offloading and non-offloading states in Defnition \ref{Def:Of_NOf}. In this subsection, we mathematically identify the conditions for a state to be an offloading and non-offloading state for a given time horizon. This is formally stated in Proposition \ref{Pro:ONO_conds}.
	\begin{proposition}\label{Pro:ONO_conds}
		\textit{Assuming that two states $\mathbf{s}$, $\mathbf{s}_a \in \mathbb{S}_{adj}\left(\mathbf{s}\right)$, and a time horizon $T$ is given. $\mathbf{s}_a$ is a non-offloading state if and only if the following inequality holds}
		\begin{align}
			J_T\left(\mathbf{s}_a\right) - J_T\left(\mathbf{s}\right) < C_o.
		\end{align}
		\textit{Otherwise, $\mathbf{s}_a$ is an offloading state}.
	\end{proposition}
	\textit{Proof}: Please see Appendix \ref{proof_Pro:ONO_conds}.
	
	Assume we are given a state $\mathbf{s}$ and a time horizon $T$ with $L^*$ as the optimal offloading decision. We recall that $\bar{\mathbf{s}}_{1L}$ is the  state that resulted by removing the first $L$ most imminent tasks from $\mathbf{s}$. Then, the next property of the optimal offloading policy is stated as follows.
	\begin{theorem}\label{Theo:L*_thesmallest}
		\textit{The optimal offloading decision is the smallest offloading decision $L$ such that $\bar{\mathbf{s}}_L$ is a non-offloading state}.
	\end{theorem}
	\textit{Proof}: Please see Appendix \ref{proof_Theo:L*_thesmallest}.

	\section{Optimal Offloading Policy} \label{Sec:OptimalPolicy}
	\subsection{Optimal Policy Description}
	\ifbulletlist
	{\color{red}
	\begin{enumerate}
	    \item Description of the optimal policy implementation.
	\end{enumerate}
	}
	\fi
	
	\iftext{
	
	Given an initial state $\mathbf{s}$ and a time horizon $T$, whenever local processing is available to process a task, the most imminent task will be processed. When a AMA is present,  the optimal policy consists of two steps. 
	
	\textbf{Step 1.} Tasks are offloaded from $\mathbf{s}$ following Algorithm \ref{Gen2Red_Algo} to reach a reduced state $\mathbf{s}_{r}$. Moreover the value of $L_g$, the number of offloaded tasks is determined.
	
	\textbf{Step 2.} In this step, the DP equation is solved recursively and  
	\[L_r = {\argmin_{L \in \mathbb{L}\left(\mathbf{s}_r\right)}}\left\{\mathcal{C}\left(\mathbf{s}_r, L\right) + G_{T-1}\left(\mathbf{s}_r, L\right) \right\}\]
	tasks are offloaded from $\mathbf{s}_r$. Note that as described in equation (\ref{eq:gen2sem_eq}), the recursion involves evaluation of only a finite number states in each one of the terms $J_{T-1}\left(\cdot\right), J_{T-2}\left(\cdot\right)$ etc. for any state $\mathbf{s}$. The optimal number of tasks that are offloaded from $\mathbf{s}$ is $L^* = L_g + L_r $.

	The computational load required to solve the recursive DP equation (\ref{DP_Eq}) can be further reduced by the following strategy. Every time the optimal decision $L_r$ associated with a reduced state $\mathbf{s}_r$ is computed for a given $T$, the triplet $\left(\mathbf{s}_r, L^*, T\right)$ is saved. This can be done for all the reduced states, as the number of reduced states is finite. Then, for every $\mathbf{s}_r$ obtained from Step 1 for a given $T$, the corresponding $L^*$ can be retrieved instantly.

	}
	\fi

	\ifbulletlist
	{\color{red}
	\begin{enumerate}
	    \item How memory usage is reduced for the optimal policy described above.
	\end{enumerate}
	}
	\fi
	
	\iftext{
	
	By exploiting the properties presented in Theorem \ref{Theo:adjacent_L*}, the computational burden can be further reduced as follows. Let us consider a sequence of adjacent states: $\mathbf{s}_1, \ldots, \mathbf{s}_i, \ldots$ 
  in which $\mathbf{s}_{i+1} \in \mathbb{S}_{adj}\left(\mathbf{s}_i\right)$. Assume the optimal decision of state $\mathbf{s}_i, i \ge 1$ is known, and denoted by $L^*_i$. From Theorem \ref{Theo:adjacent_L*}, the optimal decisions $L^*_{i-u}$ of states $\mathbf{s}_{i-u}, u = 1, \ldots, i-1$, can be inferred as follows
	\begin{align}
		L^*_{i-u} &= \max\left(L^*_i - u, 0 \right), u = 1, \ldots, i-1. \label{eq:tr_down}
	\end{align}
	In the case when $L^*_i \ge 1$ for state $\mathbf{s}_i$, the optimal decision for states $\mathbf{s}_{i+v}, v = 1, 2, \ldots$ are computed by
	\begin{align}
	L^*_{i+v} &= L^*_i + v, v = 1, 2 \ldots    \label{eq:tr_up}
	\end{align}
	In general, the optimal offloading decisions of all the states $\mathbf{s}_{i+v}$ and $\mathbf{s}_{i-u}$ mentioned above can be obtained without relying on the DP equation (\ref{DP_Eq}).
	
	In the next section, we provide numerical results to verify the results presented in this paper thus far.
	}
	\fi
	
	\section{Numerical Results} \label{Sec:NumericalResults}
	\ifbulletlist
	{\color{red}
	\begin{enumerate}
	    \item Introduction to numerical results section.
	\end{enumerate}
	}
	\fi
	
	\iftext{
	In this section, we present numerical examples that help visualize some of the properties and equations as well as show the memory savings of the numerical computations of the DP equation (\ref{DP_Eq}) with the aid of Eq. (\ref{eq:gen2sem_eq}) described in Sec. \ref{sec:semi-reduced}. Some of the examples utilize different sets of parameters to illustrate the performance of the system under different parameter configurations.
	
	\begin{table}
		\centering
		\begin{tabular}{c|c|c|c|c|c|c|c}
			  $T$   & $p_u$ & $C_p$ & $C_o$ & $\mu$ & $p_0$ & $N$ & $p_i \;\forall i\colon 1\leq i\leq N$\\ \hline
			$1,000$ & $0.7$ & $ 3 $ & $ 1 $ & $0.7$ & $0.5$ & $3$ & $1/6$ \\
		\end{tabular}
		\caption{System Parameters for Section \ref{seC:opt-off-d}}
		\label{tbl:sys-param-lin-eq}
	\end{table}
	}
	\fi
	
	\subsection{Optimal Offloading Decision Visualization} \label{seC:opt-off-d}
	\ifbulletlist
	{\color{red}
	\begin{enumerate}
	    \item Numerical verification of adjacent states.
	\end{enumerate}
	}
	\fi
	
	\iftext{
	In this example, we illustrate the idea presented in Theorem \ref{Theo:L*_thesmallest} visually for a system with the dimension of state vector to be $N=3$, i.e. $\mathbf{s} = (n_1,n_2,n_3)$. We represent states as coordinates in a 3-D state space. Then, for $n_3=0,1,2,3$, we consider the 2-D slices of this space and depict them as Figs.~\ref{fig:n30}-\ref{fig:n33} correspondingly.
    We use the system parameters described in Table \ref{tbl:sys-param-lin-eq}. The cases when $n_3 \ge 3$ would result in similar figures as Fig.~\ref{fig:n33} except that the optimal offloading decision for each offloading state would increase by $n_3 - 2$. In these figures, red dots represent offloading states, and black dots represent non-offloading states.
    From states with component $n_2 \geq 1$ in Fig. \ref{fig:n30}, such as $\left(0, 2, 0\right)$, $\left(1, 2, 0\right)$, $\left(1, 1, 0\right)$, $\left(2, 1, 0\right)$, etc., we can reach the non-offloading state $\left(0, 1, 0\right)$ with a smaller number of offloaded tasks  than state $\left(0, 0, 0\right)$.
    A similar argument applies for states with component $n_2 = 0$, like $\left(1, 0, 0\right)$, $\left(2, 0, 0\right)$, etc.,  whose ``nearest'' non-offloading state is  $\left(0, 0, 0\right)$.
    
    In Fig. \ref{fig:n31}, only the state $\left(0, 0, 1\right)$ is non-offloading. The optimal offloading decisions of all the offloading states shown are the smallest number of most imminent tasks to be offloaded to reach state $\left(0, 0, 1\right)$.
    Fig. \ref{fig:n33} does not have any non-offloading state. For example, the optimal offloading decision for the state $\left(0, 0, 3\right)$ is 1 to reach the non-offloading state $\left(0, 0, 2\right)$ which is shown in Fig. \ref{fig:n31}.
    \begin{figure}
	    \begin{subfigure}[b]{\textwidth}
		    \centering
		    \includegraphics[width=0.5\textwidth]{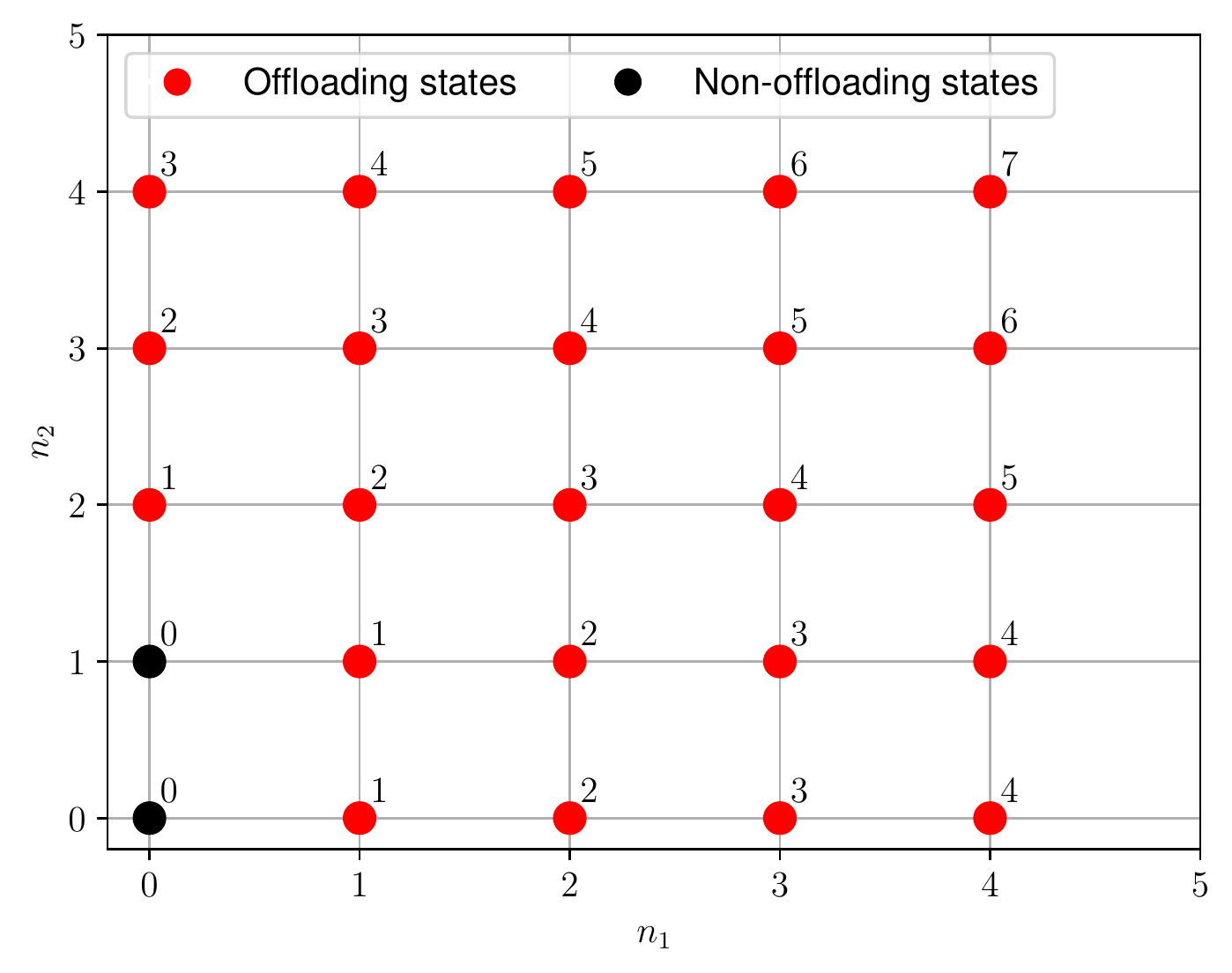}
		    \caption{$n_3 = 0$}
		    \label{fig:n30}
	    \end{subfigure}
	    \hfill
    	\begin{subfigure}[b]{\textwidth}
    		\centering
	    	\includegraphics[width=0.5\textwidth]{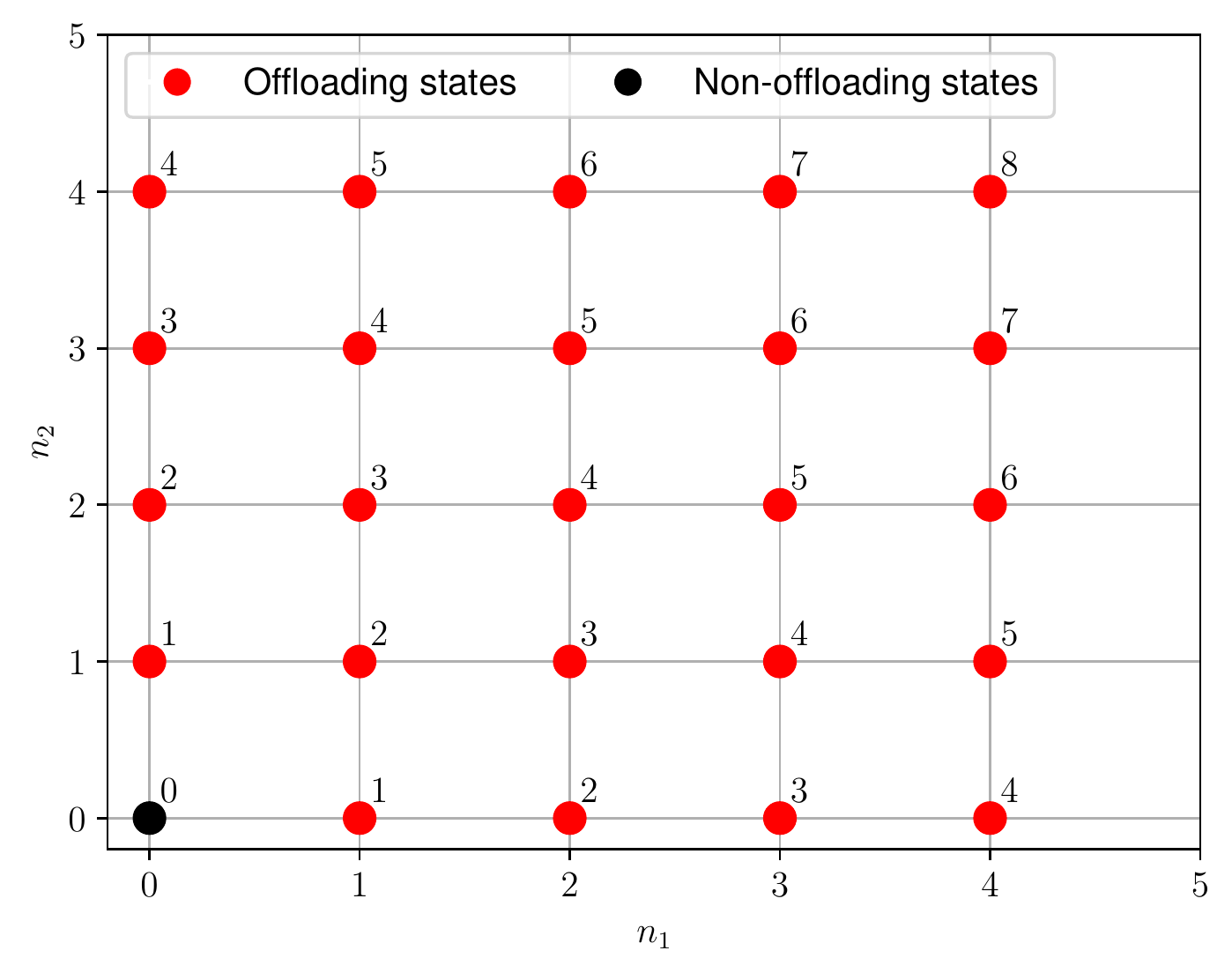}
		    \caption{$n_3 = 1$}
		    \label{fig:n31}
	    \end{subfigure}
	    \hfill
    	\begin{subfigure}[b]{\textwidth}
    		\centering
    		\includegraphics[width=0.5\textwidth]{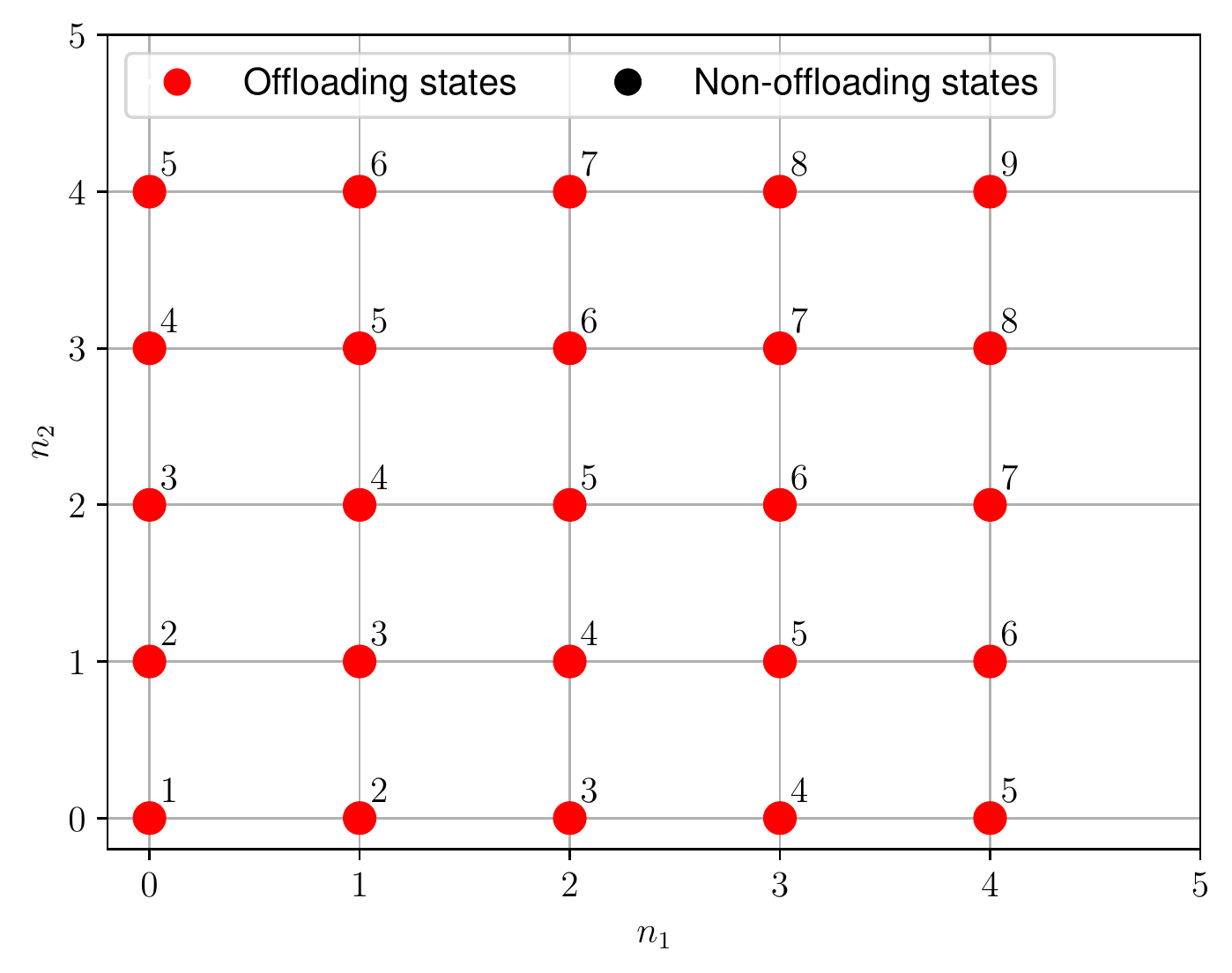}
    		\caption{$n_3 = 2$}
    		\label{fig:n33}
    	\end{subfigure}
    	\caption{Offloading states for varying $n_3$ values}
	\end{figure}
	}
	\fi

	\subsection{Optimal Offloading Decisions for Adjacent States}
	\ifbulletlist
	{\color{red}
	\begin{enumerate}
	    \item Numerical verification of adjacent states.
	\end{enumerate}
	}
	\fi
	
	\iftext{
    In Fig.~\ref{fig:convex-dp-n}-\ref{fig:convex-dp-l} we visualize the results of equations (\ref{eq:tr_down}), (\ref{eq:tr_up}) and Theorem \ref{Theo:adjacent_L*}. The figures show the optimal offloading decisions marked with a star symbol in adjacent states. For these examples, all figures have a state size $N=5$ over a time horizon $T = 1,000$. Fig.~\ref{fig:convex-dp-n}, \ref{fig:convex-dp-o} and \ref{fig:convex-dp-l} use the parameters listed in Tables \ref{tbl:opt-off-n}, \ref{tbl:opt-off-o} and \ref{tbl:opt-off-l}, respectively. On the vertical axis, we graph the minimum cost $\mathcal{C}^{\text{A}}\left(\mathbf{s}_i, L\right) + G_{T}^{\text{A}}\left(\mathbf{s}_i, L\right)$ attained by offloading $L$ most imminent tasks from state $\mathbf{s}_i$ given that the AMA is available. This is introduced in Eq. (\ref{eq:JsL_def}). 
    
    In these three figures, for $i = 1, 2, 3, 4$, we consider the states   $\mathbf{s}_1 = \left(0, 0, 0, 0, 1\right)$, $\mathbf{s}_2 = \left(0, 0, 0, 0, 2\right)$, $\mathbf{s}_3 = \left(0, 0, 1, 0, 2\right)$, and $\mathbf{s}_4 = \left(0, 1, 1, 0, 2\right)$. These states are chosen such that $\mathbf{s}_{i}$ is adjacent to $\mathbf{s}_{i+1}, i = 1, 2, 3$. The presented results indicate that the optimal offloading decision of a state differs from that of its adjacent state by 1, or both are capped at 0, as  Eqs. (\ref{eq:tr_down})-(\ref{eq:tr_up}) and Theorem \ref{Theo:adjacent_L*} suggest. For example, in Fig. \ref{fig:convex-dp-n}, the optimal decision of $\mathbf{s}_4$ is 2, and that of $\mathbf{s}_3$ is 1, hence, the difference is 1. The same observation applies for the pair $\mathbf{s}_3$ and $\mathbf{s}_2$. The optimal decision of $\mathbf{s}_2$ is 0, therefore, that of $\mathbf{s}_1$ is also 0. Figs~\ref{fig:convex-dp-o} and \ref{fig:convex-dp-l} present the same properties. 

	\begin{table}
		\centering
		\begin{tabular}{c|c|c|c|c|c|c}
			$p_u$ & $C_p$ & $C_o$ & $\mu$ & $p_0$ & $p_i \; \forall i\colon 1\leq i\leq N$ \\ \hline
			$0.5$ & $ 3 $ & $ 1 $ & $0.5$ & $0.5$ & $0.1$ \\
		\end{tabular}
		\caption{System Parameters for Fig. \ref{fig:convex-dp-n}}
		\label{tbl:opt-off-n}
	\end{table}
	
	\begin{table}
		\centering
		\begin{tabular}{c|c|c|c|c|c|c}
			$p_u$ & $C_p$ & $C_o$ & $\mu$ & $p_0$ & $p_i \; \forall i\colon 1\leq i\leq N$\\ \hline
			$0.4$ & $ 4 $ & $ 1 $ & $0.3$ & $0.3$ & $0.14$\\
		\end{tabular}
		\caption{System Parameters for Fig. \ref{fig:convex-dp-o}}
		\label{tbl:opt-off-o}
	\end{table}
	
	\begin{table}
		\centering
		\begin{tabular}{c|c|c|c|c|c|c}
			$p_u$ & $C_p$ & $C_o$ & $\mu$ & $p_0$ & $p_i \; \forall i\colon 1\leq i\leq N$\\ \hline
			$0.7$ & $ 2 $ & $ 1 $ & $0.6$ & $0.6$ & $0.08$\\
		\end{tabular}
		\caption{System Parameters for Fig. \ref{fig:convex-dp-l}}
		\label{tbl:opt-off-l}
	\end{table}

    \begin{figure}
    \centering
	\begin{subfigure}[b]{\textwidth}
		\centering
		\includegraphics[width=0.5\textwidth]{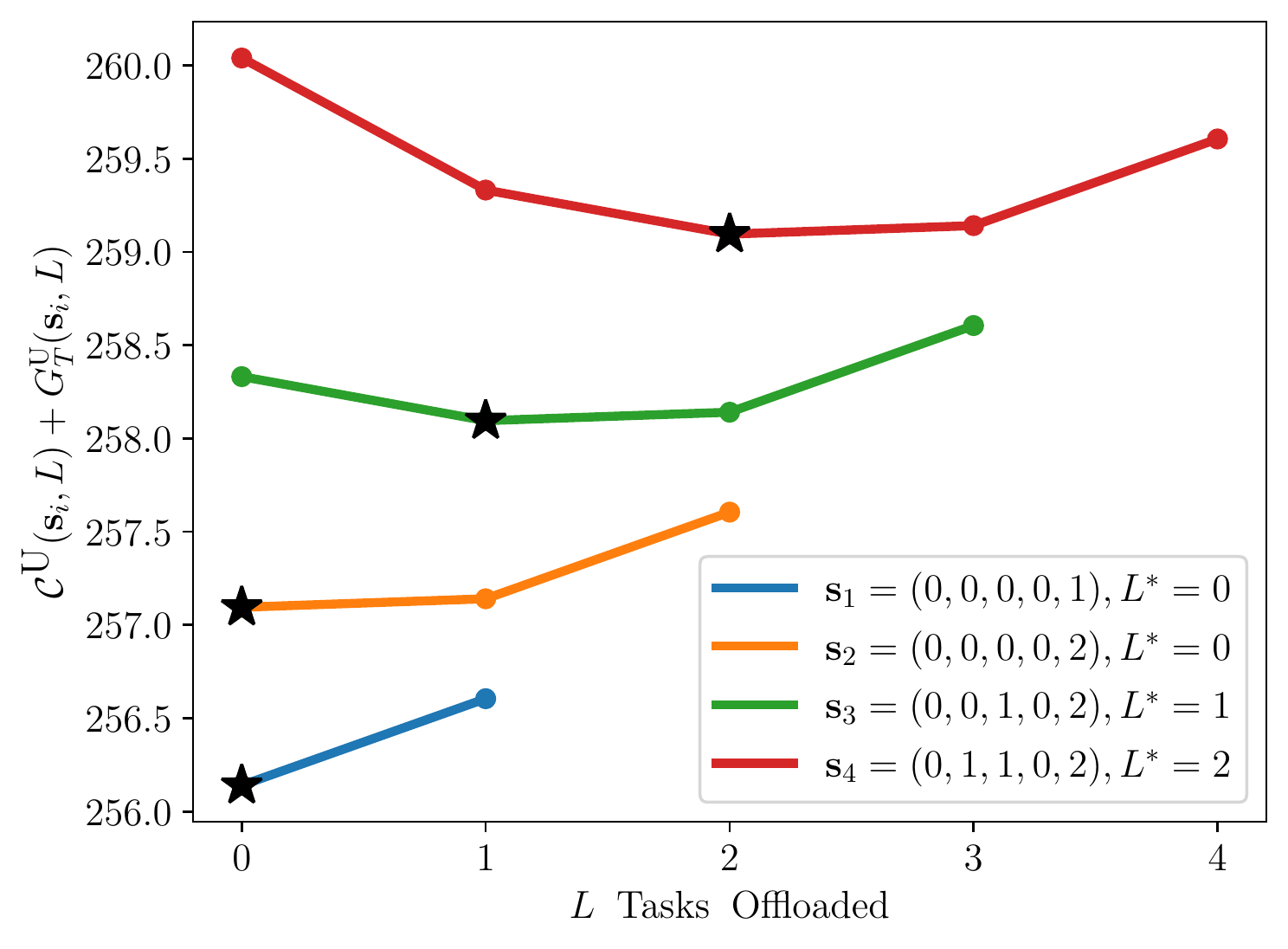}
		\caption{Adjacent states using parameters in Table \ref{tbl:opt-off-n}}
		\label{fig:convex-dp-n}
	\end{subfigure}
	
	\begin{subfigure}[b]{\textwidth}
		\centering
		\includegraphics[width=0.5\textwidth]{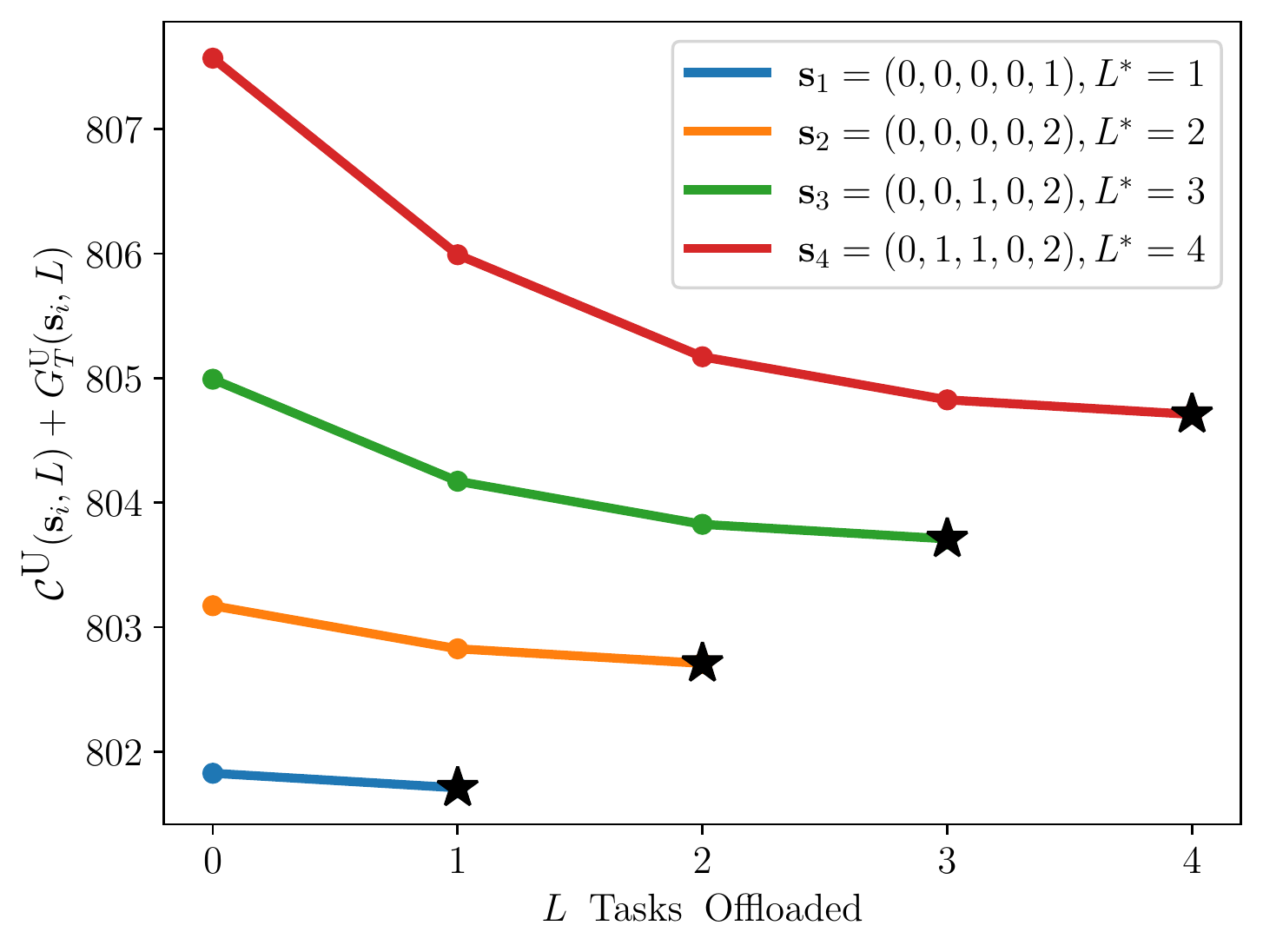}
		\caption{Adjacent states using parameters in Table \ref{tbl:opt-off-o}}
		\label{fig:convex-dp-o}
	\end{subfigure}
	
	\begin{subfigure}[b]{\textwidth}
		\centering
		\includegraphics[width=0.5\textwidth]{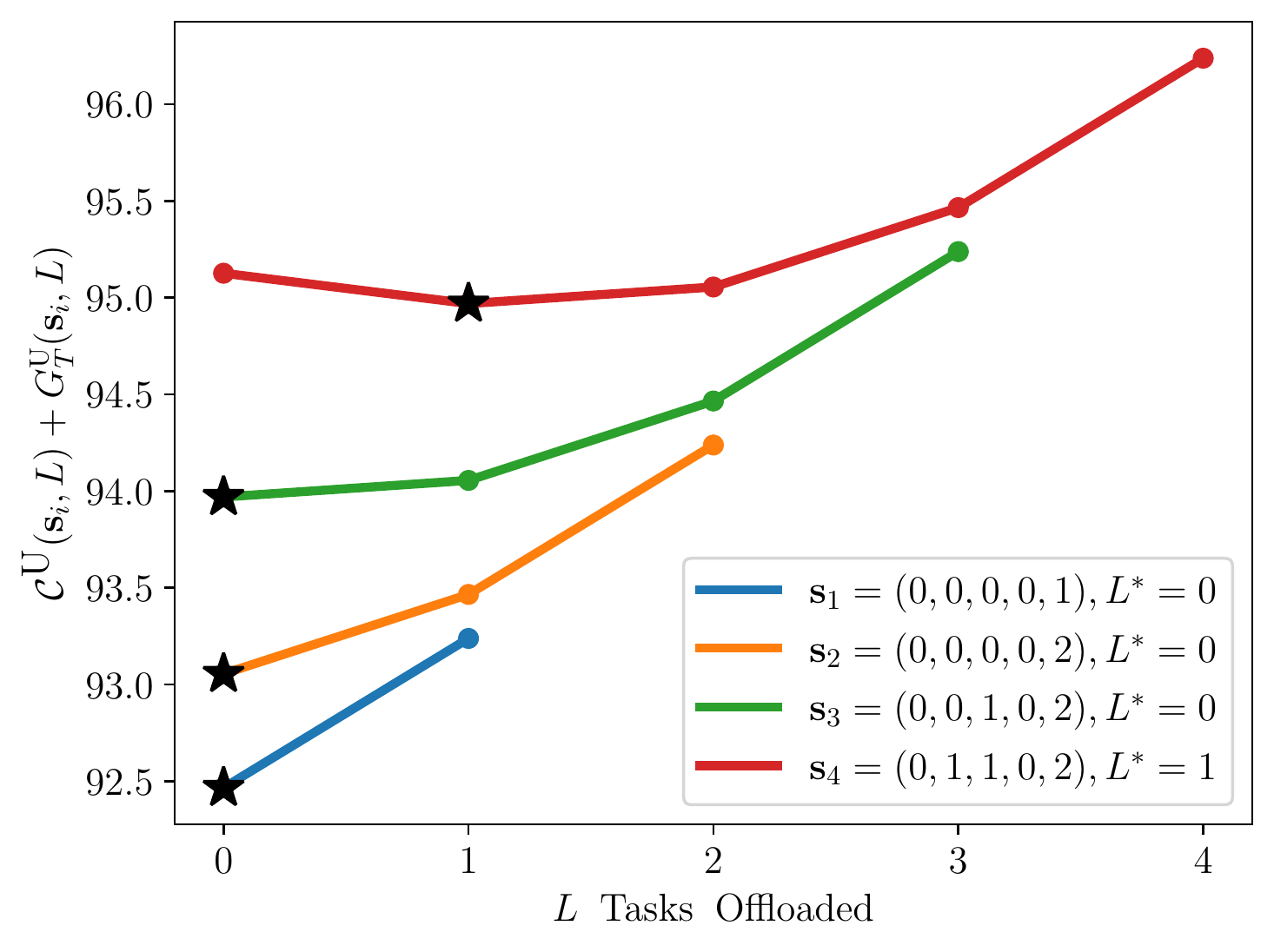}
		\caption{Adjacent states using parameters in Table \ref{tbl:opt-off-l}}
		\label{fig:convex-dp-l}
	\end{subfigure}
	\caption{Visual interpretation of Eqs. (\ref{eq:tr_down})-(\ref{eq:tr_up}) and Theorem \ref{Theo:adjacent_L*}.}
	\end{figure}
	}
	\fi
	
	\subsection{Memory Savings Using Equation (\ref{eq:gen2sem_eq})}
	\ifbulletlist
	{\color{red}
	\begin{enumerate}
	    \item Memory saving by utilizing the lean result.
	\end{enumerate}
	}
	\fi
	
	\iftext{
    In order to numerically compute the DP equation (\ref{DP_Eq}),
	we store the computed value of $J_T(\mathbf{s})$ in memory for a given arbitrary state $\mathbf{s}$ and time horizon $T$.
	By using Eq. (\ref{eq:gen2sem_eq}), the size of memory required is reduced.  This is because saving $J_T(\cdot)$ values is only required for lean states $\mathbf{s}_m$ where the number of lean states is smaller than that of ``generic'' states. Note that these savings are achieved at the expense of calculating the term $C_{g2m}$ in Eq.  (\ref{eq:gen2sem_eq}). 
	
	In Fig.~\ref{fig:memory-saving-n5}, \ref{fig:memory-saving-offload-n4}, and Fig.~\ref{fig:memory-saving-process-n3}, we show the resulting difference in the number
	of $J_T(\cdot)$ values
	saved for 2 cases when $N=5$, $N=4$ and $N=3$, respectively. Note that different parameters other than the system parameters $N$ and $T$ will not affect the memory savings. In the first case, Eq. (\ref{DP_Eq}) is used, while in the second case, Eq. (\ref{DP_Eq}) is used with the aid of Eq. (\ref{eq:gen2sem_eq}).
	The line in blue represents the values saved using only Eq. (\ref{DP_Eq}), while the line in red
	utilizes both Eqs. (\ref{DP_Eq}) and (\ref{eq:gen2sem_eq}).
	
	\begin{figure}
	\begin{subfigure}[b]{\textwidth}
		\centering
		\includegraphics[width=0.5\textwidth]{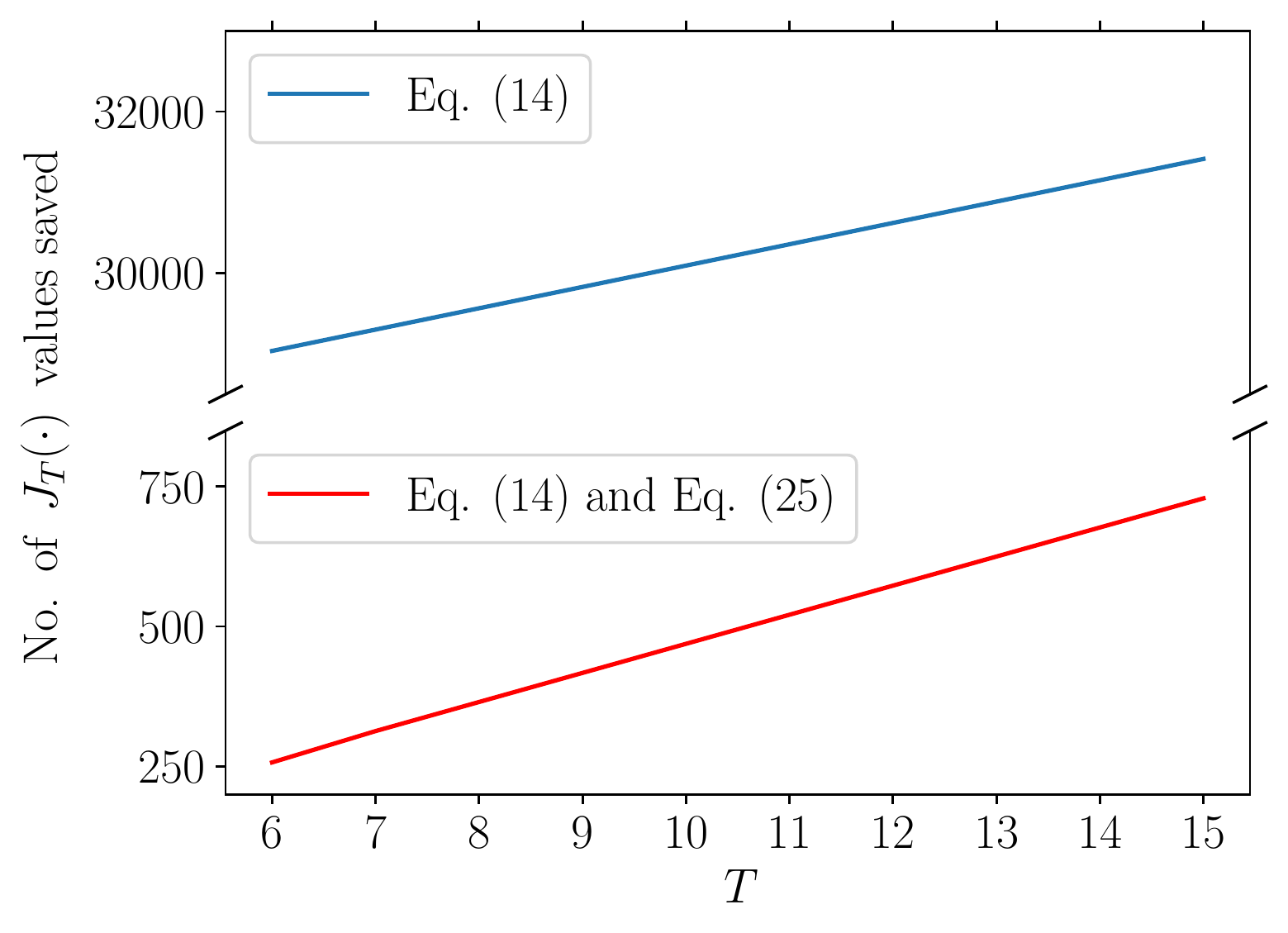}
		\caption{Memory savings with $N=5$}
		\label{fig:memory-saving-n5}
	\end{subfigure}

	\begin{subfigure}[b]{\textwidth}
		\centering
		\includegraphics[width=0.5\textwidth]{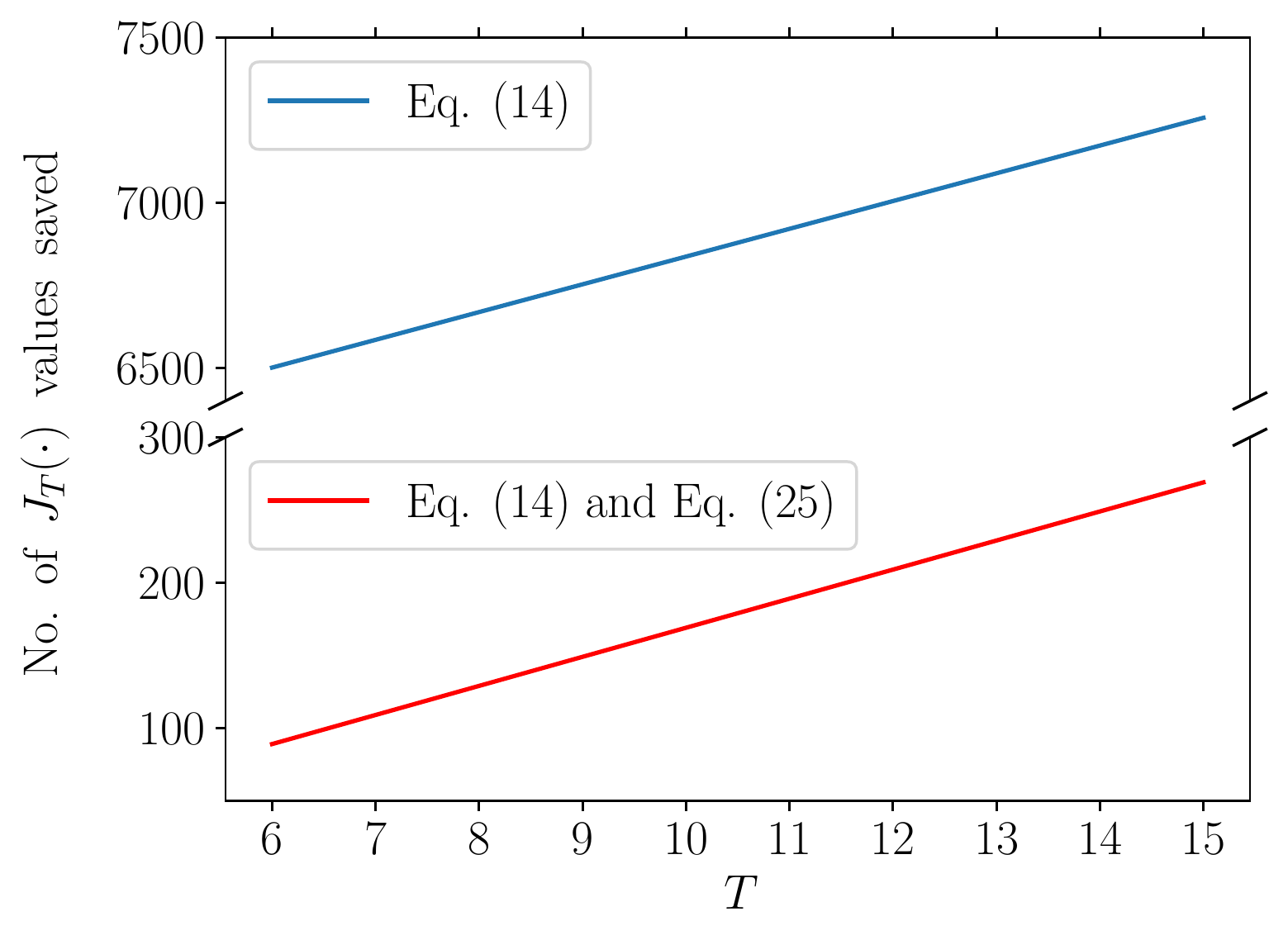}
		\caption{Memory Saving with $N=4$}
		\label{fig:memory-saving-offload-n4}
	\end{subfigure}

	\begin{subfigure}[b]{\textwidth}
		\centering
		\includegraphics[width=0.5\textwidth]{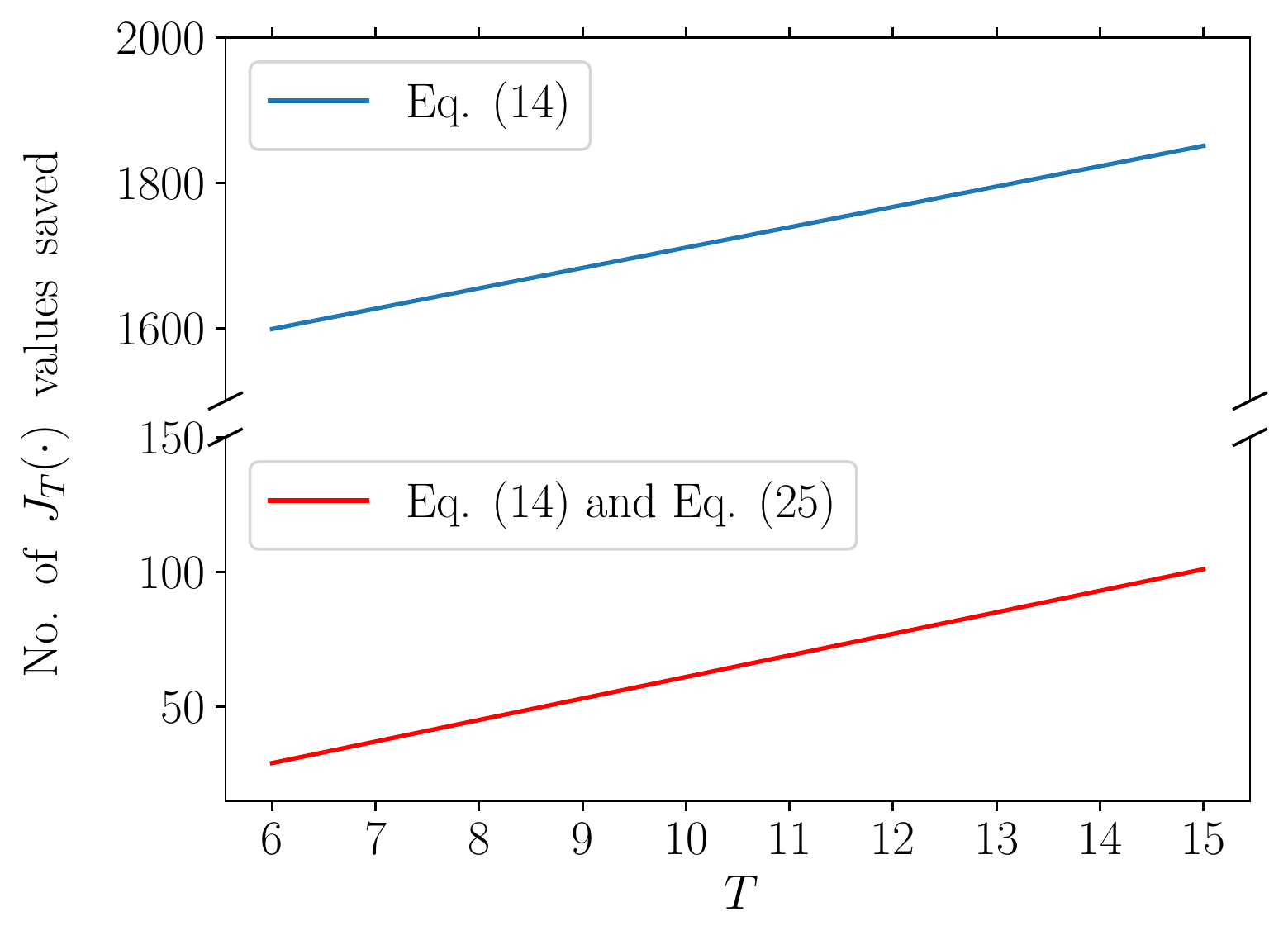}
		\caption{Memory Saving with $N=3$}
		\label{fig:memory-saving-process-n3}
	\end{subfigure}
	\caption{Memory Savings using Eq. (\ref{eq:gen2sem_eq})}
	\end{figure}
    }
	\fi

    \section{Conclusion} \label{Sec:Conclusion}
	In this work, we studied a mobile edge computing system with dynamic user demand. In the context of an optimal stochastic control framework for serving user tasks, we  considered the following features: tasks with firm deadlines, their random offloading to a remote server (AMA) or their processing by a local server (BS) with intermittent service. We considered an expected time-average cost over a finite time horizon and formulated a Dynamic Programming problem towards the minimization of this cost. In order to tackle the ``Curse of Dimensionality'', we studied important characteristics of the optimal policy and reduced the computational load for its calculation. In particular, we proved that the DP equation can be evaluated for every given state (in the infinite state space of our model) by considering a specific finite space called lean state space. Further reduction in the computational load was achieved by using the concept of ``adjacent states''. This allowed us to evaluate the optimal cost for all such states from knowledge of the cost in only one state. Finally, 
	based on these properties, we described an optimal task offloading  policy.
	
	\section{Appendix A: Proofs of Theorems} \label{Sec:AppendixTheorems}
 \comment{Hereafter, we will introduce some new notations to facilitate the presentation of our proofs in this section and the subsequent two sections.} For a given initial state $\mathbf{s}$, and a time horizon $T$, we recall that $\bar{\mathbf{s}}_{1L}$ is the state obtained by offloading $L$ most imminent tasks from state $\mathbf{s}$. Combining this definition with that of $J_T^{\text{A}}\left(\mathbf{s}, L\right)$ in Eq. (\ref{eq:JsL_def}), we have the following relation
	\begin{align}\label{rel_orig}
		J_T^{\text{A}}\left(\mathbf{s}, L\right) = J_T^{\text{A}}\left(\bar{\mathbf{s}}_{1L}, 0\right) + LC_o, \text{ for } L \in \mathbb{L}\left(\mathbf{s}\right)
	\end{align}
	which can be explained as follows. We note that by offloading $L$ most imminent tasks from $\mathbf{s}$, we pay a cost $LC_o$, and arrive at state $\bar{\mathbf{s}}_{1L}$. Therefore, if we wish to describe the offloading of $L$ tasks on the left hand side of Eq. (\ref{rel_orig}), this is equivalent to removing the $L$ most imminent tasks from state $\mathbf{s}$ to reach state $\bar{\mathbf{s}}_{1L}$, and offloading 0 task from $\bar{\mathbf{s}}_{1L}$. Finally, we further add the offloading cost $LC_o$ to the overall cost. The right hand side of Eq. (\ref{rel_orig}) describes this idea.
	
	Since the minimum average cost when offloading no task is the same regardless of the presence of the AMA's availability, i.e.,
	\begin{align}\label{equi_noAMA}
		J_T^{\text{A}}\left(\bar{\mathbf{s}}_{1L}, 0\right) = J_T^{\overline{\text{A}}}\left(\bar{\mathbf{s}}_{1L}\right), \text{ for } L \in \mathbb{L}\left(\mathbf{s}\right),
	\end{align}
	we have
	\begin{align}\label{JsL_intro}
		J_T^{\text{A}}\left(\mathbf{s}, L\right) = J_T^{\overline{\text{A}}}\left(\bar{\mathbf{s}}_{1L}\right) + LC_o, \text{ for } L \in \mathbb{L}\left(\mathbf{s}\right).
	\end{align}
	
	Furthermore, considering two offloading decisions $L_1$ and $L_2$ where $L_1+L_2, L_1, L_2 \in \mathbb{L}\left(\mathbf{s}\right)$, we note that offloading $L_1 + L_2$ tasks from the original state $\mathbf{s}$ would result in the same state that is obtained by offloading $L_2$ tasks from $\bar{\mathbf{s}}_{1L_1}$. An example for this point is as follows:
    \begin{example}
        Assume $\mathbf{s} = \left( 0, 1, 2, 2, 3\right), L_1 = 2, L_2 = 3$, then, the resulting states by offloading $L_1 + L_2 = 5$ and $L_1 = 2$ tasks from $\mathbf{s}$, respectively, are
        \begin{align}
            & \bar{\mathbf{s}}_{1\left(L_1 + L_2\right)} = \left(0, 0, 0, 0, 3 \right), \label{eq:example_fresh} \\
            & \bar{\mathbf{s}}_{1L_1} = \left(0, 0, 1, 2, 3 \right).
        \end{align}
        Now, offloading  $L_2 = 3$ tasks from $\bar{\mathbf{s}}_{1L_1}$ would give us the same state as in Eq. (\ref{eq:example_fresh}).
    \end{example}
    From Eq. (\ref{JsL_intro}), we have
    \begin{align}
        J^{\text{A}}_T\left(\mathbf{s}, L_1 + L_2 \right) &= J^{\overline{\text{A}}}_T\left(\bar{\mathbf{s}}_{1\left(L_1 + L_2\right)}\right) + \left(L_1 + L_2\right)C_o, \\
        J^{\text{A}}_T\left(\bar{\mathbf{s}}_{1L_1}, L_2 \right) &= J^{\overline{\text{A}}}_T\left(\bar{\mathbf{s}}_{1\left(L_1 + L_2\right)}\right) + L_2C_o.
    \end{align}
    The above two expressions result in the following one:
    \comment{
    We can describe the minimum average cost attained by offloading $L_1+L_2$ tasks from $\mathbf{s}$ by the following two components: 
	\begin{enumerate}
		\item The cost $L_1C_o$ to offload $L_1$ tasks from $\mathbf{s}$ and arrive at state $\bar{\mathbf{s}}_{1L_1}$.
		\item The minimum average cost attained by offloading $L_2$ tasks from $\bar{\mathbf{s}}_{1L_1}$, i.e., $J_T^{\text{A}}\left(\bar{\mathbf{s}}_{1L_1}, L_2\right)$.
	\end{enumerate}
	Therefore, we have the following expression
    }
	\begin{align}\label{JsL_intro1}
		J_T^{\text{A}}\left(\mathbf{s}, L_1+L_2\right) = J_T^{\text{A}}\left(\bar{\mathbf{s}}_{1L_1}, L_2\right) + L_1C_o.
	\end{align}

	\subsection{Proof of Theorem \ref{Theo:adjacent_L*}}\label{proof_Theo:adjacent_L*}
    
    From Lemma \ref{Lem:Convexity}, we have that function $F^{\overline{\text{A}}}\left(T, \mathbf{s}, d, L\right)$ is discrete convex with respect to $L$ for every given $T$, $d$, and $\mathbf{s} = \left(n_1, \ldots, n_N\right)$. Let us consider the case when $d=1$, and we call $L^* \in \mathbb{L}_1\left(\mathbf{s}\right)$ the value at which $F^{\overline{\text{A}}}\left(T, \mathbf{s}, 1, L\right)$ attains its minimum. From Lemma \ref{Lem:Convex2L*}, $L^*$ is also the optimal offloading decision for $\mathbf{s}$. Hence,
	\begin{align} \label{optim_at_s}
		J_T^{\text{A}}\left(\mathbf{s}, L\right) \ge J_T^{\text{A}}\left(\mathbf{s}, L^*\right), \text{ for all } L \in \mathbb{L}_1\left(\mathbf{s}\right),
	\end{align}
	
	Let us consider the following cases:
	\begin{itemize}
		\item If $L^* \ge 1$ which means that $\mathbf{s} \not\equiv \left(0, \ldots, 0\right)$,  the set of inequalities (\ref{optim_at_s}) can be re-written as
		\begin{align}\label{ineq:convex2L*_1}
			J_T^{\text{A}}\left(\mathbf{s}, 1 + L-1\right) \ge J_T^{\text{A}}\left(\mathbf{s}, 1 + L^*-1\right), \text{ for all } L \in \mathbb{L}_1\left(\mathbf{s}\right) \backslash \left\{0\right\}.
		\end{align}
		Applying Eq. (\ref{JsL_intro1}) with $L_1=1$, $L_2 = L-1$ to the left-hand side of Ineqs. (\ref{ineq:convex2L*_1}), and with $L_2=L^*-1$ to the right-hand side of Ineqs. (\ref{ineq:convex2L*_1}), we have
		\begin{align}\label{optim_s11}
			J_T^{\text{A}}\left(\bar{\mathbf{s}}_{11}, L-1\right) \ge J_T^{\text{A}}\left(\bar{\mathbf{s}}_{11}, L^*-1\right), \text{ for all } L, L^* \in \mathbb{L}_1\left(\mathbf{s}\right) \backslash \left\{0\right\},
		\end{align}
		 in which we recall that $\bar{\mathbf{s}}_{11}$ is obtained by offloading the most imminent task from $\mathbf{s}$, thus, $\mathbf{s} \in \mathbb{S}_{adj}\left(\bar{\mathbf{s}}_{11}\right)$. 

		 Since $L, L^* \in \mathbb{L}_1\left(\mathbf{s}\right) \backslash \left\{0\right\}$, we have  $L-1, L^*-1 \in \mathbb{L}_1\left(\bar{\mathbf{s}}_{11}\right)$ \footnote{Note: The set $\mathbb{L}_1\left(\mathbf{s}\right)$ is the same as $\mathbb{L}_1(\bar{\mathbf{s}}_{11})$ except that $\mathbb{L}_1\left(\mathbf{s}\right)$ contains an element which is the total number of tasks in $\mathbf{s}$, while $\mathbb{L}_1(\bar{\mathbf{s}}_{11})$ does not contain that element.}. Therefore, Ineqs. (\ref{optim_s11}) suggests that $L^*-1$ is the optimal offloading decision for $\bar{\mathbf{s}}_{11}$. This proves the first point of Theorem \ref{Theo:adjacent_L*}.
		
		\item If $L^*=0$, i.e., $\mathbf{s}$ is a non-offloading state. Then, if $\mathbf{s}$ has only one task, we have $\bar{\mathbf{s}}_{11} \equiv \left(0, \ldots, 0\right)$. Therefore, $\bar{\mathbf{s}}_{11}$ is a non-offloading state trivially.
		
		If $\mathbf{s}$ has at least 2 tasks. From the convexity of the function $F^{\overline{\text{A}}}\left(T, \mathbf{s}, d, L\right) = J_T^{\overline{\text{A}}}\left(\bar{\mathbf{s}}_{dL}\right) + LC_o$, and the condition that $L^*=0$, we have
		\begin{align} \label{ineq:convex&L*=0}
			\begin{split}
			    &J_T^{\overline{\text{A}}}\left(\mathbf{s}\right) \le J_T^{\overline{\text{A}}}\left(\bar{\mathbf{s}}_{11}\right) + C_o, \text{ for all } L \in \mathbb{L}_1\left(\mathbf{s}\right)\backslash \left\{0, 1\right\},\\
                &J_T^{\overline{\text{A}}}\left(\bar{\mathbf{s}}_{11}\right) + C_o \le J_T^{\overline{\text{A}}}\left(\bar{\mathbf{s}}_{1L}\right) + LC_o, \text{ for all } L \in \mathbb{L}_1\left(\mathbf{s}\right)\backslash \left\{0, 1\right\}.
			\end{split}
		\end{align}
  where the two inequalities in Ineq. (\ref{ineq:convex&L*=0}) are from the optimality of $L^*$, and the convexity of cost functions proven in Subsec. \ref{proof_Lem:Convexity}, respectively.
  
		Applying Eq. (\ref{equi_noAMA}) to the second inequality of Ineqs. (\ref{ineq:convex&L*=0}), we have
		\begin{align} \label{optim_L*=0}
			J_T^{\text{A}}\left(\bar{\mathbf{s}}_{11}, 0\right) + C_o \le J_T^{\text{A}}\left(\bar{\mathbf{s}}_{1L}, 0\right) + LC_o, \text{ for all } L \in \mathbb{L}_1\left(\mathbf{s}\right)\backslash \left\{0, 1\right\}.
		\end{align}
		Using Eq. (\ref{JsL_intro1}) with $L_1=1$ and $L_2 = L-1$ gives us
		\begin{align}\label{mani1}
			J_T^{\text{A}}\left(\mathbf{s}, L\right) = J_T^{\text{A}}\left(\bar{\mathbf{s}}_{11}, L-1\right) + C_o.
		\end{align}
		Also, using Eq. (\ref{JsL_intro1}) with $L_1=L$ and $L_2 = 0$ yields
		\begin{align}\label{mani2}
			J_T^{\text{A}}\left(\mathbf{s}, L\right) = J_T^{\text{A}}\left(\bar{\mathbf{s}}_{1L}, 0\right) + LC_o.
		\end{align}
		From Eqs. (\ref{mani1}) and (\ref{mani2}), we have
		\begin{align}\label{mani2bis}
			J_T^{\text{A}}\left(\bar{\mathbf{s}}_{11}, L-1\right) + C_o = J_T^{\text{A}}\left(\bar{\mathbf{s}}_{1L}, 0\right) + LC_o.
		\end{align}
		Combining Eq. (\ref{mani2bis}) with Ineqs. (\ref{optim_L*=0}) gives the following set of inequalities:
		\begin{align} 
			J_T^{\text{A}}\left(\bar{\mathbf{s}}_{11}, 0\right) \le J_T^{\text{A}}\left(\bar{\mathbf{s}}_{11}, L-1\right), \text{ for all } L \in \mathbb{L}_1\left(\mathbf{s}\right)\backslash \left\{0, 1\right\}.
		\end{align}
		By replacing $L-1$ with $\tilde{L}$ in the above inequalities, we have
		\begin{align} 
			J_T^{\text{A}}\left(\bar{\mathbf{s}}_{11}, 0\right) \le J_T^{\text{A}}\left(\bar{\mathbf{s}}_{11}, \tilde{L}\right), \text{ for all } \tilde{L} \in \mathbb{L}_1\left(\bar{\mathbf{s}}_{11}\right)\backslash \left\{0\right\},
		\end{align}
		indicating that $\bar{\mathbf{s}}_{11}$ is a non-offloading state. This proves the second point of Theorem \ref{Theo:adjacent_L*}.
	\end{itemize}
	
	Finally, from the first two points of Theorem \ref{Theo:adjacent_L*}, we can conclude that, for given time horizon $T$, if $L^* \ge 1$ is the optimal offloading decision for state $\mathbf{s}$, $L^*_a = L^* + 1$ is optimal for every state $\mathbf{s}_a \in \mathbb{S}_{adj}\left(\mathbf{s}\right)$. This is because if $L^*_a \ne L^* + 1$ and $L^*_a \ge 1$, from the first point of Theorem \ref{Theo:adjacent_L*}, the optimal decision for $\mathbf{s}$ must be $L^*_a - 1 \ne L^*$ which is a contradiction. Moreover, if $L^*_a = 0$, from the second point of Theorem \ref{Theo:adjacent_L*}, $\mathbf{s}$ must be a non-offloading state, leading to another contradiction. This proves the third point of Theorem \ref{Theo:adjacent_L*}.
	
	\subsection{Proof of Theorem \ref{Theo:L*_thesmallest}}\label{proof_Theo:L*_thesmallest}

    We recall that given an current state $\mathbf{s}$, $\bar{\mathbf{s}}_{1L}$ denotes the resulting state by offloading $L$ most imminent task from $\mathbf{s}$. If $L^* = 0$ is the optimal offloading decision of state $\mathbf{s}$, then, $\mathbf{s}$ is a non-offloading state. It is trivially that $L^* = 0$ is the smallest offloading decision to reach a non-offloading state.
    
    Let us consider the sequence of states $\bar{\mathbf{s}}_{10} = \mathbf{s}, \bar{\mathbf{s}}_{11}, \bar{\mathbf{s}}_{12}, \ldots, \bar{\mathbf{s}}_{1i}, \ldots$. By definition of the notation $\bar{\mathbf{s}}_{1i}$, state $\bar{\mathbf{s}}_{1\left(i+1\right)}$ is obtained by offloading the most imminent task from state $\bar{\mathbf{s}}_{1i}$ in the sequence. Therefore, state $\bar{\mathbf{s}}_{1i}$ is adjacent to $\bar{\mathbf{s}}_{1\left(i+1\right)}$. Assume $L^* > 0$ is the optimal offloading decision for $\mathbf{s}$. From the first point of Theorem \ref{Theo:adjacent_L*}, the optimal offloading decision of state $\bar{\mathbf{s}}_{11}$ would be $L^*_{1}=L^*-1$. By alternatively applying this property, the optimal offloading decisions $L^*_i$ of states $\bar{\mathbf{s}}_{1i}$ for $i=1, \ldots, L^*$ can be derived as
    \begin{align}
        L^*_i = L^* - i, \text{ for } i = 1, \ldots, L^*.
    \end{align}

    The above result suggests that the optimal offloading decision of the state $\bar{\mathbf{s}}_{1L^*}$ would be $L^*_{i} = 0$ for $i=L^*$. From the second point of Theorem \ref{Theo:adjacent_L*}, the optimal offloading decision of $\bar{\mathbf{s}}_{1\left(L^*+1\right)}$ would also be 0. Repeatedly applying this property allows us to derive the optimal decisions for state $\bar{\mathbf{s}}_{1i}, i > L^*$ as follows:
    \begin{align}
        L^*_i = 0, \text{ for } i > L^*.
    \end{align}
    
    In conclusion, states $\bar{\mathbf{s}}_{1i}$ for $i \le L^* - 1$ are offloading states, and states $\bar{\mathbf{s}}_{1i}$ for $i \ge L^*$ are non-offloading states. Therefore, $L^* > 0$ is the smallest offloading decision to reach a non-offloading state $\bar{\mathbf{s}}_{1L^*}$.

    \comment{
    \textcolor{red}{Recall:}
    
    \textcolor{red}{$F^{\overline{\text{A}}}\left(T, \mathbf{s}, d, L \right)$ is the minimum average cost that can be attained if, at the current time slot, we are forced to offload L tasks starting from deadline $d$ from state $\mathbf{s}$.}

    \textcolor{red}{$F^{\overline{\text{A}}}\left(T, \mathbf{s}, d, L \right)$ is convex wrt. $L$. Assume $F^{\overline{\text{A}}}\left(T, \mathbf{s}, d, L \right)$ attains its minimum at $L^*$ for $d=1$, then, $L^*$ is the optimal offloading decision associated with state $\mathbf{s}$ and time horizon $T$.}
 
	For a given offloading state $\mathbf{s} = \left(n_1, \ldots, n_N\right)$, and a time horizon $T$, we assume that the function $F^{\overline{\text{A}}}\left(T, \mathbf{s}, d, L \right) = J^{\overline{\text{A}}}\left(\bar{\mathbf{s}}_{d L}\right) + LC_o$ attains its minimum at $L^*$ for $d=1$. From Lemma \ref{Lem:Convex2L*}, $L^*$ is the optimal offloading decision for the given initial state $\mathbf{s}$ and time horizon $T$. We have the following set of inequalities:
	\begin{align}\label{ineq:rep_FL*}
	J^{\overline{\text{A}}}\left(\bar{\mathbf{s}}_{1 L^*}\right) + L^* C_o \le J^{\overline{\text{A}}}\left(\bar{\mathbf{s}}_{1 L}\right) + LC_o, \text{ for all } L \in \mathbb{L}_1\left(\mathbf{s}\right).
	\end{align}
	If $L^* = \sum_{i=1}^N n_i$, it is trivially that $\bar{\mathbf{s}}_{1L^*} \equiv \left(0, \ldots, 0\right)$ is a non-offloading state. If $L^* < \sum_{i=1}^N n_i$, from Ineq. (\ref{ineq:rep_FL*}), we have
	\begin{align}
		J_T^{\overline{\text{A}}}\left(\bar{\mathbf{s}}_{1L^*}\right) + L^*C_o \le J_T^{\overline{\text{A}}}\left(\bar{\mathbf{s}}_{1\left(L^*+1\right)}\right) + \left(L^*+1\right)C_o,
	\end{align}
	or equivalently,
	\begin{align}
		J_T^{\overline{\text{A}}}\left(\bar{\mathbf{s}}_{1L^*}\right) \le J_T^{\overline{\text{A}}}\left(\bar{\mathbf{s}}_{1\left(L^*+1\right)}\right) + C_o,
	\end{align}
	where $\bar{\mathbf{s}}_{1L^*} \in \mathbb{S}_{adj}\left(\bar{\mathbf{s}}_{1\left(L^*+1\right)}\right)$. This is because $\bar{\mathbf{s}}_{1L^*}$ is obtained by offloading $L^*$ most imminent tasks from $\mathbf{s}$, while $\bar{\mathbf{s}}_{1\left(L^*+1\right)}$ is obtained by offloading $L^*+1$ most imminent tasks from $\mathbf{s}$. According to Proposition \ref{Pro:ONO_conds}, $\bar{\mathbf{s}}_{1L^*}$ is a non-offloading state.
	
	As $F^{\overline{\text{A}}}\left(T, \mathbf{s}, d, L\right)$ is discrete convex with respect to $L$, and attains the minimum at $L^*$, we have
	\begin{align}
		J_T^{\overline{\text{A}}}\left(\bar{\mathbf{s}}_{1\left(L^*-L\right)}\right) + \left(L^*-L\right)C_o \le J_T^{\overline{\text{A}}}\left(\bar{\mathbf{s}}_{1\left(L^*-L-1\right)}\right) + \left(L^*-L-1\right)C_o, ~ L = 0, \ldots, L^*-1.
	\end{align}
	We note that as $\mathbf{s}$ is an offloading state, $L^* \ge 1$. The above inequality is equivalent to
	\begin{align}\label{ineq_done}
		J_T^{\overline{\text{A}}}\left(\bar{\mathbf{s}}_{1\left(L^*-L\right)}\right) + C_o \le J_T^{\overline{\text{A}}}\left(\bar{\mathbf{s}}_{1\left(L^*-L-1\right)}\right), ~ L = 0, \ldots, L^*-1,
	\end{align}
	in which $\bar{\mathbf{s}}_{1\left(L^*-L-1\right)} \in \mathbb{S}_{adj}\left(\bar{\mathbf{s}}_{1\left(L^*-L\right)}\right)$. Based on Proposition \ref{Pro:ONO_conds}, the set of inequalities (\ref{ineq_done}) suggest that states $\bar{\mathbf{s}}_{1\left(L^*-L-1\right)}$ associated with $L = 0, \ldots, L^*-1$ are all offloading states, where $\bar{\mathbf{s}}_{1\left(L^*-L-1\right)} \equiv \mathbf{s}$ when $L = L^*-1$. This leads to the conclusion that $L^*$ is the smallest offloading decision for $\mathbf{s}$ to obtain a non-offloading state $\bar{\mathbf{s}}_{1L^*}$.
    }
 
	\section{Appendix B: Proofs of Lemmas}\label{Sec:AppendixLemmas}
	\subsection{Proof of Lemma \ref{lem_Catalan}}\label{proof_lem_Catalan}
	A sequence $(a_1,a_2, \ldots, a_N)$ of non-negative integers is called {\em Catalan} if 
	\begin{align}
		\label{eq:catalan}
		1 \leq a_1 \leq a_2 \leq \cdots \leq a_N \mbox{ and } a_i \leq i, \mbox{ for all $1 \leq i \leq N$.}
	\end{align}
	
	In the proof we show that there is a one-to-one correspondence betweeb reduced sequences and Catalan sequences of the same length $N$. The correspondence is defined as follows.
	
	Given a reduced sequence $(n_1,n_2, \ldots, n_N)$, which satisfies inequalities \eqref{ineq:reduced_cond}, define a sequence $(a_1,a_2, \ldots, a_N)$ as follows 
	$$
	a_i =  (n_1+1) +n_2+\cdots+n_i  .
	$$
	It is clear that the resulting sequence satisfies the Catalan sequence property~\eqref{eq:catalan}. 
	
	Conversely, given a Catalan sequence $(a_1,a_2, \ldots, a_N)$, which satisfies property~\eqref{eq:catalan}, define the sequence $(n_1,n_2, \ldots, n_N)$ as follows
	$$
	n_i = \left\{
	\begin{array}{ll}
		0 & \mbox{ if $i=1$} \\
		a_i - a_{i-1} & \mbox{ if $2 \leq i \leq N$}
	\end{array}
	\right.
	$$
	The resulting sequence contains elements of a reduced state vector because 
	\begin{align*}
		n_1 + n_2 + n_3+ \cdots + n_i 
		&=  0 + a_2 -a_1 + a_3-a_2 + \cdots + a_i - a_{i-1}\\
		&= a_i -a_1 \\
		&\leq i-1 ,
	\end{align*}
	since $a_1 = 1$. Also observe that the resulting correspondence between reduced and Catalan sequences of the same length $N$ is one-to-one.
	
	The proof of this lemma is now complete since in exercise 78 from \cite{stanley2015catalan}, the number of Catalan sequences of length $N$ is equal to the Catalan number $C_N$.
	
	\subsection{Proof of Lemma \ref{Lem:MI_optimality}} \label{proof_Lem:MI_optimality}
	We assume that an offloading state $\mathbf{s} = \left(n_1, \ldots, n_N\right)$ is given with $d$ is the deadline of the most imminent task. It is trivially that when $\mathbf{s} = \left(0, \ldots, 0\right)$, the optimal decision is offloading 0 task. When $N \ge 2$, in the following two cases, the optimal decision is offloading the most imminent task:
	\begin{itemize}
	    \item $d \le N-1$, and $n_i=0$ for $i = d+1, \ldots, N$.
	    \item $d=N$.
	\end{itemize}
	Examples for these cases are given below:
	\begin{example}
	In the following two examples, it is trivially that the optimal policy offloads the most imminent tasks:
	\begin{itemize}
	\item $\mathbf{s} = \left(0, 0, 4, 0, 0\right)$ where $N = 5$, $d = 3$, and $n_i=0, i = 4, 5$.
	\item $\mathbf{s} = \left(0, 0, 0, 0, 7\right)$ where $d = N = 5$.
	\end{itemize}
	\end{example}
	
	Now, we will consider the remaining case which is: $N \ge 2$, and $d \le N-1$, and there exists a deadline $d' \in \left\{d+1, \ldots, N \right\}$ such that $n_{d'} > 0$. This indicates that $n_i = 0, i = 1, \ldots, d-1$, $n_d \ge 1$, and $n_{d'} \ge 1$. Let us recall the following notation:
	\begin{itemize}
	\item $\bar{\mathbf{s}}_{d1}$ is the state obtained by offloading a task at deadline $d$ from state $\mathbf{s}$.
	\item $\bar{\mathbf{s}}_{d'1}$ is the state obtained by offloading a task at deadline $d'$ from state $\mathbf{s}$.
	\end{itemize}
	We notice that the $i^{\text{th}}$ elements of state $\bar{\mathbf{s}}_{d1}$ is defined by
	\begin{align}
	\begin{cases}
	n_d - 1, &\text{ if } i = d, \\
	n_i, & \text{ if } i \ne d.
	\end{cases}
	\end{align}
	The $i^{\text{th}}$ elements of state $\bar{\mathbf{s}}_{d'1}$ is defined by
	\begin{align}
	\begin{cases}
	n_{d'} - 1, & \text{ if } i = d', \\
	n_i, &\text{ if } i \ne d'.
	\end{cases}
	\end{align}
	Therefore, we have: $\bar{\mathbf{s}}_{d1} \in \mathbb{S}_{pp}\left(\bar{\mathbf{s}}_{d'1}\right)$. From Proposition \ref{Pro:deadline_postponed}, we have
	\begin{align}
	J_T\left(\bar{\mathbf{s}}_{d1}\right) \le J_T\left(\bar{\mathbf{s}}_{d'1}\right), \text{ for every given time horizon } T.
	\end{align}
	This indicates that state $\bar{\mathbf{s}}_{d1}$ is associated with a lower average cost than that of state $\bar{\mathbf{s}}_{d'1}$ for every deadline $d' \in \left\{d+1, \ldots, N \right\}$. Therefore, if $\mathbf{s}$ is an offloading state, it is optimal to offload the most imminent task from $\mathbf{s}$. Moreover, from the above proof, we can also conclude that whenever the local processing is available, it is optimal to process the most imminent task in $\mathbf{s}$.
	
	To this end, by considering state $\bar{\mathbf{s}}_{d1}$ in the place of state $\mathbf{s}$ and repeating the proof above, we have that: If $\bar{\mathbf{s}}_{d1}$ is an offloading state, it is optimal to offload the most imminent task from $\bar{\mathbf{s}}_{d1}$. Therefore, we can conclude that: If $\mathbf{s}$ is an offloading state and the corresponding optimal offloading decision is 2, it is optimal to offload 2 most imminent tasks from $\mathbf{s}$. Keeping repeating the same analysis leads us to the final conclusion as follows: If $\mathbf{s}$ is an offloading state, and the corresponding optimal offloading decision is $L^*$, it is optimal to offload the $L^*$ most imminent tasks from $\mathbf{s}$.
	
	\subsection{Proof of Lemma \ref{Lem:Convexity}}\label{proof_Lem:Convexity}
	From the definition of function $F^{\overline{\text{A}}}\left(T, \mathbf{s}, d, L \right)$ in Eq. (\ref{F_def}), we notice that $LC_o$ is discrete linear, and hence, discrete convex with respect to $L$. Therefore, in order to prove that $F^{\overline{\text{A}}}\left(T, \mathbf{s}, d, L \right)$ is discrete convex with respect to $L$, we need to prove that
	\begin{align}\label{FbarU_remind}
	f^{\overline{\text{A}}}\left(T, \mathbf{s}, d, L \right) = J_T^{\overline{\text{A}}}\left(\bar{\mathbf{s}}_{dL} \right)
	\end{align}
	is discrete convex with respect to $L$. To begin with, let us define two other functions as follows:
	\begin{align}\label{F_def_remind}
		f\left(T, \mathbf{s}, d, L \right) = J_T\left(\bar{\mathbf{s}}_{dL}\right), \\
		f^{\text{A}}\left(T, \mathbf{s}, d, L \right) = J_T^{\text{A}}\left(\bar{\mathbf{s}}_{dL}\right). \label{FU_remind}
	\end{align}
	The three functions $f\left(T, \mathbf{s}, d, L \right)$, $f^{\text{A}}\left(T, \mathbf{s}, d, L \right)$, and $f^{\overline{\text{A}}}\left(T, \mathbf{s}, d, L \right)$ are characterized by the parameters $T$, $\mathbf{s}$, and $d$. All of them take $L$ as variable. The domains of these three functions are the same as that of $F^{\overline{\text{A}}}\left(T, \mathbf{s}, d, L \right)$ which are given in Eqs. (\ref{F_domain1}) and (\ref{F_domain2}).
	
	The definitions of $J^{\text{A}}_T\left(\cdot\right)$ and $J^{\overline{\text{A}}}_T\left(\cdot\right)$ are given in Eqs. (\ref{eq:JAMA}) and (\ref{eq:JNoAMA}). In this subsection, we re-express $J^{\text{A}}_T\left(\bar{\mathbf{s}}_{dL} \right)$ and $J^{\overline{\text{A}}}_T\left(\bar{\mathbf{s}}_{dL} \right)$ as follows:
	\begin{align}
			&J^{\text{A}}_T\left(\bar{\mathbf{s}}_{dL} \right) = \underset{L' \in \mathbb{L}\left(\bar{\mathbf{s}}_{dL} \right)}{\min} \left\{\mathcal{C}^{\text{A}}\left(\bar{\mathbf{s}}_{dL} , L'\right) + G^{\text{A}}\left(\bar{\mathbf{s}}_{dL} , L'\right)\right\}, \\
			&J^{\overline{\text{A}}}_T\left(\bar{\mathbf{s}}_{dL} \right) = \mathcal{C}^{\overline{\text{A}}}\left(\bar{\mathbf{s}}_{dL}\right) + G^{\overline{\text{A}}}\left(\bar{\mathbf{s}}_{dL}\right) = \mathcal{C}^{\text{A}}\left(\bar{\mathbf{s}}_{dL} , 0\right) + G^{\text{A}}\left(\bar{\mathbf{s}}_{dL} , 0\right).
	\end{align}

	By denoting $M_d = \sum_{i=d}^N n_i$. We can express the functions $f^{\text{A}}\left(T, \mathbf{s}, d, L \right)$ and $f^{\overline{\text{A}}}\left(T, \mathbf{s}, d, L \right)$ as follows:
	\begin{align}
		\underset{L \in \mathbb{L}_d\left(\mathbf{s}\right)}{f^{\text{A}}\left(T, \mathbf{s}, d, L \right)} & = \underset{\hat{L} \in \left\{0, \ldots, M_d-L\right\}}\min\left\{\left(L+\hat{L}\right)C_o +  G_{T-1}\left(\mathbf{s}, d, L+\hat{L} \right) \right\} + \mathds{1}_{d \ge 2}n_1C_p, \label{F_asmin} \\
		\underset{L \in \mathbb{L}_d\left(\mathbf{s}\right)}{f^{\overline{\text{A}}}\left(T, \mathbf{s}, d, L \right)} & =  G_{T-1}\left(\mathbf{s}, d, L\right) + \mathds{1}_{d \ge 2} n_1 C_p. \label{F_noAMA}
	\end{align}
	
	In (\ref{F_asmin}), we denote $\bar{L} = L + \hat{L}$. Then, the average future cost function $G_T\left(\mathbf{s}, d, \bar{L} = L+\hat{L}\right)$ is given by
	\begin{align}\label{G_new}
		G_{T-1}\left(\mathbf{s}, d,  \bar{L}\right) = \mu \sum_{k=0}^N p_k J_{T-1}\left(\mathbf{s}_{d \bar{L} k}^{'} \right) + \left(1-\mu\right) \sum_{k=0}^N p_k J_{T-1}\left(\mathbf{s}_{d \bar{L} k}^{''} \right)
	\end{align}
	where $\mathbf{s}_{d \bar{L} k}^{'}$ is the state transited from $\mathbf{s}$ with the following steps: offloading $\bar{L} = L+\hat{L}$ most imminent tasks starting from deadline $d$, deadline shifting, a task arrives with deadline $k$ ($k=0$ implies no task arrival), and local processing. $\mathbf{s}_{d \bar{L} k}^{''}$ is defined similarly but without the local processing at the end.
	
	Subsequently, we will show that each term $J_{T-1}\left(\cdot\right)$ on the right-hand side of Eq. (\ref{G_new}) can be presented by the family of function $f$. Let us consider an initial state $\mathbf{s} = \left(n_1, \ldots, n_N\right) \not\equiv \left(0, \ldots, 0\right)$. First of all, we need to find states $\mathbf{s}^1$ and $\mathbf{s}^2$ such that $\mathbf{s}^1_{d \bar{L}} \equiv \mathbf{s}_{d \bar{L} k}'$ and $\mathbf{s}^2_{d \bar{L}} \equiv \mathbf{s}_{d \bar{L} k}''$  where $\mathbf{s}^1_{d \bar{L}}$ and $\mathbf{s}^2_{d \bar{L}}$ are obtained by removing $\bar{L}$ most imminent tasks having deadline greater than or equal to $d$ from $\mathbf{s}^1$ and $\mathbf{s}^2$, respectively. Then, we will have the followings:
	\begin{align}
		f\left(T-1, \mathbf{s}^1, d', L'\right) &= J_{T-1}\left(\bar{\mathbf{s}}^1_{d' L'}\right) = J_{T-1}\left(\mathbf{s}_{d \bar{L} k}'\right), \label{eq:J_to_f_goal1}\\
		f\left(T-1, \mathbf{s}^2, d'', L''\right) &= J_{T-1}\left(\bar{\mathbf{s}}^2_{d'' L''}\right) = J_{T-1}\left(\mathbf{s}_{d \bar{L} k}''\right). \label{eq:J_to_f_goal2}
	\end{align}
	Hereafter, we will defined $\mathbf{s}^1$, $\mathbf{s}^2$, $d'$, $d''$, $L'$, and $L''$ in different cases, and provide examples to support interpretation.
\begin{itemize}
	\item When $k+1 < d$ which implies $d \ge 2$, we have: $\mathbf{s}^1 \equiv \mathbf{s}'_k$, $\mathbf{s}^2 \equiv \mathbf{s}''_k$, $d' = d'' = d-1$, and $L' = L'' = \bar{L}$. The state $\mathbf{s}'_k$ is obtained from $\mathbf{s}$ by: performing deadline shifting,  adding a new task with deadline $k$ ($k=0$ implies no new task is added), and removing the most imminent task. State $\mathbf{s}''_k$ is defined similarly as state $\mathbf{s}'_k$, but without removing the most imminent task at the end. 

\begin{example}
As an example for the case ``$k + 1 < d$'', we consider an initial state $\mathbf{s} = \left(0, 6, 0, 5, 7 \right)$, and a time horizon $T >  5$. We assume that $d=3$, and $\bar{L}=7$ most imminent tasks having deadline greater than or equal to $d$ are offloaded, and $k = 1$ implying that a new task arrives with deadline 1 in the next time slot. Then, we have
\begin{align}
\mathbf{s}'_{d \bar{L} k} = \mathbf{s}'_{371} = \left(6, 0, 0, 5, 0 \right), \label{eq:exCase1_1}\\
\mathbf{s}''_{d \bar{L} k} = \mathbf{s}''_{371} = \left(7, 0, 0, 5, 0 \right). \label{eq:exCase1_2}
\end{align} 
The states $\mathbf{s}'_k$ and $\mathbf{s}''_k$ in this example are
\begin{align}
\mathbf{s}'_k = \mathbf{s}'_1 = \left(6, 0, 5, 7, 0\right), \label{eq:exCase1_1b} \\
\mathbf{s}''_k = \mathbf{s}'_1 = \left(7, 0, 5, 7, 0\right). \label{eq:exCase1_2b}
\end{align}
With $\mathbf{s}^1 \equiv \mathbf{s}'_k$, $\mathbf{s}^2 \equiv \mathbf{s}''_k$, $d' = d'' = d-1 = 2$, and $L' = L'' = \bar{L} = 7$ as presented above, the states $\bar{\mathbf{s}}^1_{d' L'}$ and $\bar{\mathbf{s}}^2_{d'' L''}$ are given by
\begin{align}
\bar{\mathbf{s}}^1_{d' L'} = \mathbf{s}^1_{2 7} = \left(6, 0, 0, 5, 0 \right), \label{eq:exCase1_3} \\
\bar{\mathbf{s}}^2_{d'' L''} = \mathbf{s}^2_{2 7} = \left(7, 0, 0, 5, 0 \right). \label{eq:exCase1_4}
\end{align}
This shows that state $\bar{\mathbf{s}}^1_{d' L'}$ in Eq. (\ref{eq:exCase1_3}) is the same as state $\mathbf{s}'_{d \bar{L} k}$ in Eq. (\ref{eq:exCase1_1}). Also, state $\bar{\mathbf{s}}^2_{d'' L''}$ in Eq. (\ref{eq:exCase1_4}) is the same as state $\mathbf{s}''_{d \bar{L} k}$ in Eq. (\ref{eq:exCase1_2}). Therefore, Eqs. (\ref{eq:J_to_f_goal1}) and (\ref{eq:J_to_f_goal2}) can be achieved.
\end{example}

\item When $k+1 \ge d$, let $\beta_{dk} = \sum_{i=d}^{k+1} n_i$, we have:  
\begin{itemize}
\item[--] If $d \ge 2$: For $\bar{L} \le \beta_{dk}$, $\mathbf{s}^1 \equiv \mathbf{s}'_k$, $\mathbf{s}^2 \equiv \mathbf{s}''_k$, $d' = d'' = d-1$, and $L' = L'' = \bar{L}$. For $\bar{L} > \beta_{dk}$, $\mathbf{s}^1 \equiv \mathbf{s}'_{d \beta_{dk} k}$, $\mathbf{s}^2 \equiv \mathbf{s}''_{d \beta_{dk} k}$, $d' = d'' = k+1$, and $L' = L'' = \bar{L}- \beta_{dk}$.
\item[--] If $d = 1$: For $\bar{L} \le \beta_{dk}$, $\mathbf{s}^1 \equiv \mathbf{s}'_k$, $\mathbf{s}^2 \equiv \mathbf{s}''_k$, $d' = d'' = d$, and $L' = L'' = \bar{L}-n_1$. For $\bar{L} > \beta_{dk}$, $\mathbf{s}^1 \equiv \mathbf{s}'_{d \beta_{dk} k}$, $\mathbf{s}^2 \equiv \mathbf{s}''_{d \beta_{dk} k}$, $d' = d'' = k+1$, and $L' = L'' = \bar{L}- \beta_{dk}$.
\end{itemize}
We note that state $\mathbf{s}'_{d \beta_{dk} k}$ is obtained following the steps: offloading $\beta_{dk}$ most imminent tasks having deadline greater than or equal to $d$ from the initial state $\mathbf{s}$, performing deadline shifting, adding a task with deadline $k$, and removing the most imminen task. State $\mathbf{s}''_{d \beta_{dk} k}$ is defined in the same way but without removing the most imminent task at the end. 

\begin{example}
Two examples for this case corresponding to $d \ge 2$ and $d=1$ are as follows. Let us consider an initial state $\mathbf{s} = \left(0, 3, 0, 7, 5\right)$, a time horizon $T > 5$, and $\bar{L}=9$. For the first example, we assume that $d=2$, and $k=2$. We have $\beta_{dk} = \beta_{22} = 3$. The states $\mathbf{s}'_{d \bar{L} k}$ and $\mathbf{s}''_{d \bar{L} k}$ are computed as
\begin{align}
\mathbf{s}'_{d \bar{L} k} = \mathbf{s}'_{292} = \left(0, 0, 1, 5, 0\right), \label{eq:exCase1_5}\\
\mathbf{s}''_{d \bar{L} k} = \mathbf{s}''_{292} = \left(0, 1, 1, 5, 0 \right). \label{eq:exCase1_6}
\end{align} 
Since $L = 9 > \beta_{dk} = 3$, we compute states $\mathbf{s}'_{d \beta_{dk} k}$ and $\mathbf{s}''_{d \beta_{dk} k}$ as follows:
\begin{align}
\mathbf{s}'_{d \beta_{dk} k} = \mathbf{s}'_{2 3 2} = \left(0, 0, 7, 5, 0\right), \\
\mathbf{s}''_{d \beta_{dk} k} = \mathbf{s}''_{2 3 2} = \left(0, 1, 7, 5, 0\right).
\end{align}
As presented above, we have $\mathbf{s}^1 \equiv \mathbf{s}'_k$, $\mathbf{s}^2 \equiv \mathbf{s}''_k$, $d' = d'' = k+1 = 3$, and $L' = L'' = \bar{L} - n_d= 9 - 3 = 6$. Therefore, states $\bar{\mathbf{s}}^1_{d' L'}$ and $\bar{\mathbf{s}}^2_{d'' L''}$ are computed as
\begin{align}
\bar{\mathbf{s}}^1_{d' L'} = \mathbf{s}^1_{1 9} = \left(0, 0, 1, 5, 0 \right), \label{eq:exCase1_7} \\
\bar{\mathbf{s}}^2_{d'' L''} = \mathbf{s}^2_{1 9} = \left(0, 1, 1, 5, 0 \right). \label{eq:exCase1_8}
\end{align}
Then, state $\bar{\mathbf{s}}^1_{d' L'}$ in Eq. (\ref{eq:exCase1_7}) is the same as state $\mathbf{s}'_{d \bar{L} k}$ in Eq. (\ref{eq:exCase1_5}). State $\bar{\mathbf{s}}^2_{d'' L''}$ in Eq. (\ref{eq:exCase1_8}) is the same as state $\mathbf{s}''_{d \bar{L} k}$ in Eq. (\ref{eq:exCase1_6}). Therefore, we obtained Eqs. (\ref{eq:J_to_f_goal1}) and (\ref{eq:J_to_f_goal2}).

In the second example, we assume that $d=1$ and $k = 3$.  We have $\beta_{dk} = \beta_{13} = 10$. The states $\mathbf{s}'_{d \bar{L} k}$ and $\mathbf{s}''_{d \bar{L} k}$ are
\begin{align}
\mathbf{s}'_{d \bar{L} k} = \mathbf{s}'_{193} = \left(0, 0, 0, 6, 0\right), \label{eq:exCase1_9}\\
\mathbf{s}''_{d \bar{L} k} = \mathbf{s}''_{193} = \left(0, 0, 1, 6, 0\right). \label{eq:exCase1_10}.
\end{align} 
Now $\bar{L} = 9 \le \beta_{dk} = 10$, we define state $\mathbf{s}'_k$ and $\mathbf{s}''_k$ as follows:
\begin{align}
\mathbf{s}'_k = \left(2, 0, 7, 6, 0\right), \\
\mathbf{s}''_k = \left(3, 0, 7, 6, 0 \right).
\end{align}
We have $\mathbf{s}^1 \equiv \mathbf{s}'_k$, $\mathbf{s}^2 \equiv \mathbf{s}''_k$, $d' = d'' = d = 1$, and $L' = L'' = \bar{L}-n_1 = 9$. Thus, states $\bar{\mathbf{s}}^1_{d' L'}$ and $\bar{\mathbf{s}}^2_{d'' L''}$ are computed as
\begin{align}
\bar{\mathbf{s}}^1_{d' L'} = \mathbf{s}^1_{1 9} = \left(0, 0, 0, 6, 0\right), \label{eq:exCase1_11} \\
\bar{\mathbf{s}}^2_{d'' L''} = \mathbf{s}^2_{1 9} = \left(0, 0, 1, 6, 0 \right). \label{eq:exCase1_12},
\end{align}
which suggests that state $\bar{\mathbf{s}}^1_{d' L'}$ in Eq. (\ref{eq:exCase1_11}) is the same as state $\mathbf{s}'_{d \bar{L} k}$ in Eq. (\ref{eq:exCase1_9}). State $\bar{\mathbf{s}}^2_{d'' L''}$ in Eq. (\ref{eq:exCase1_12}) is the same as state $\mathbf{s}''_{d \bar{L} k}$ in Eq. (\ref{eq:exCase1_10}). As a result, we achieve Eqs. (\ref{eq:J_to_f_goal1}) and (\ref{eq:J_to_f_goal2}).

\end{example}

\end{itemize}

From the presented cases, the terms $J_{T-1}\left(\mathbf{s}'_{d \bar{L} k}\right)$ and $J_{T-1}\left(\mathbf{s}''_{d \bar{L} k}\right)$ are expressed by functions $f$ as follows:
\begin{itemize}
\item If $k + 1 < d$:
\begin{align}
J_{T-1}\left(\mathbf{s}'_{d \bar{L} k}\right) &= f\left(T-1, \mathbf{s}'_k, d-1, \bar{L}\right), \label{eq:Jf1}\\
J_{T-1}\left(\mathbf{s}''_{d \bar{L} k}\right) &= f\left(T-1, \mathbf{s}''_k, d-1, \bar{L}\right). \label{eq:Jf2}
\end{align}

\item If $k + 1 \ge d$ and $d \ge 2$:
\begin{align}\label{eq:Jf3}
J_{T-1}\left(\mathbf{s}'_{d \bar{L} k}\right) = g'_{dk}\left(\bar{L}\right) =
\begin{cases}
f\left(T-1, \mathbf{s}'_k, d-1, \bar{L}\right) ,  & \text{ if } \bar{L} \le \beta_{dk},\\
f\left(T-1, \mathbf{s}'_{d \beta_{dk} k}, k+1, \bar{L} - \beta_{dk}\right), & \text{ if } \bar{L} > \beta_{dk},
\end{cases}
\end{align}

\begin{align}\label{eq:Jf4}
J_{T-1}\left(\mathbf{s}''_{d \bar{L} k}\right) = g''_{dk}\left(\bar{L}\right) =
\begin{cases}
f\left(T-1, \mathbf{s}''_k, d-1, \bar{L}\right) ,  & \text{ if } \bar{L} \le \beta_{dk},\\
f\left(T-1, \mathbf{s}''_{d \beta_{dk} k}, k+1, \bar{L} - \beta_{dk}\right), & \text{ if } \bar{L} > \beta_{dk},
\end{cases}
\end{align}
where $\beta_{dk} = \sum_{i=d}^{k+1} n_i$.

\item If $k + 1 \ge d$ and $d = 1$:
\begin{align}\label{eq:Jf5}
J_{T-1}\left(\mathbf{s}'_{1 \bar{L} k}\right) = g'_{1k}\left(\bar{L}\right) =
\begin{cases}
f\left(T-1, \mathbf{s}'_k, 1, \bar{L}\right) ,  & \text{ if } \bar{L} \le \beta_{1k},\\
f\left(T-1, \mathbf{s}'_{1 \beta_{1k} k}, k+1, \bar{L} - \beta_{1k}\right), & \text{ if } \bar{L} > \beta_{1k},
\end{cases}
\end{align}

\begin{align}\label{eq:Jf6}
J_{T-1}\left(\mathbf{s}''_{1 \bar{L} k}\right) = g''_{1k}\left(\bar{L}\right) =
\begin{cases}
f\left(T-1, \mathbf{s}''_k, 1, \bar{L}\right) ,  & \text{ if } \bar{L} \le \beta_{1k},\\
f\left(T-1, \mathbf{s}''_{1 \beta_{1k} k}, k+1, \bar{L} - \beta_{1k}\right), & \text{ if } \bar{L} > \beta_{1k},
\end{cases}
\end{align}
where $\beta_{1k} = \sum_{i=1}^{k+1} n_i$.
\end{itemize}
In the above, $g'_{dk}\left(\bar{L}\right)$ and $g''_{dk}\left(\bar{L}\right)$ are functions characterized by parameters $d$ and $k$, and take $\bar{L}$ as their variables.

Next, we will prove the convexity of functions $f$ using induction, starting with an initial case.
 
	\textbf{Initial case:} We consider state $\mathbf{s} = \left(n_1, \ldots, n_N\right)$, and a time horizon $T=1$. 
	
	In this case, if $N = 1$, we have $\mathbf{s} = \left(n_1\right)$, then, there is only one valid offloading decision $\mathbb{L}_1\left(\mathbf{s}\right) = \left\{n_1\right\}$. For $N \ge 2$, it is trivially that all tasks having deadline 1 are excessive tasks, and should be offloaded whenever the AMA is available, which results in a cost $n_1C_o$. If the AMA is not available at the initial time slot, tasks with deadline 1 will expire and result in a cost $n_1C_p$. Since in this case, we are considering a time horizon with only one time slot, the minimum average cost can be computed straightforwardly. Thus, we have the following
	\begin{align}
		f\left(1, \mathbf{s}, 1, L\right) &= J_1\left(\mathbf{s}_{1L}\right) = \left(p_uC_o + \left(1-p_u\right)C_p\right)n_1 + \left(L-n_1\right)C_o, \text{ for } L \in \mathbb{L}_1\left(\mathbf{s}\right), \label{eq:induct_init1} \\
		f\left(1, \mathbf{s}, d, L\right) &= J_1\left(\mathbf{s}_{dL}\right) = n_1C_p + LC_o, \text{ for } d \ge 2 \text{ and } L \in \mathbb{L}_d\left(\mathbf{s}\right). \label{eq:induct_init2}
	\end{align}
	We recall that the smallest offloading decision in the set $\mathbb{L}_1\left(\mathbf{s}\right)$ is $n_1$. From Eqs. (\ref{eq:induct_init1}) and (\ref{eq:induct_init2}), $f\left(1, \mathbf{s}, d, L\right), d = 1, \ldots, N$ are discrete linear function with respect to $L$, hence, they are discrete convex function with respect to $L$. Next is an inductive step where we prove the convexity of $f\left(T, \mathbf{s}, d, L\right)$ gven that of $f\left(T-1, \mathbf{s}, d, L\right)$ for every parameters $\mathbf{s}$ and $d$.

	\textbf{Inductive step:} Let us make the following assumption: \textit{The functions $f\left(T-1, \mathbf{s}, d, L\right)$ is discrete convex with respect to $L$ for every given state $\mathbf{s}$ and deadline $d$.}
	
	Now, we consider Eqs. (\ref{eq:Jf1})-(\ref{eq:Jf6}) with $\bar{L}$ is replaced by $L + \hat{L}$. From the above assumption, $f\left(T-1, \mathbf{s}'_k, d-1,  L + \hat{L} \right)$ in Eq. (\ref{eq:Jf1}), and $f\big(T-1, \mathbf{s}''_k, d-1, L + \hat{L} \big)$ in Eq. (\ref{eq:Jf2}) are discrete convex with respect to $L$. Next, we will prove that $g'_{dk}\left(L + \hat{L} \right)$ in Eq. (\ref{eq:Jf3}) is discrete convex with respect with respect to $L$. Then, the convexity of functions $g''_{dk}\left(L + \hat{L} \right)$ in Eq. (\ref{eq:Jf4}), $g'_{1k}\left( L + \hat{L} \right)$ in Eq. (\ref{eq:Jf5}), and $g''_{1k}\left( L + \hat{L} \right)$ in Eq. (\ref{eq:Jf6}) can also be proved in a very similar way.
	
	Let us consider function $g'_{dk}\left(L + \hat{L} \right)$ in Eq. (\ref{eq:Jf3}). With our assumption above, $f\big(T-1, \mathbf{s}'_k, d-1, L + \hat{L}\big)$ for $L + \hat{L} \le \beta_{dk}$, and $f\left(T-1, \mathbf{s}'_{d \beta_{dk} k}, k+1, L + \hat{L} - \beta_{dk}\right)$ for $L + \hat{L} > \beta_{dk}$ are discrete convex with respect to $L$. Therefore, in order to prove the convexity of $g'_{dk}\left(L + \hat{L}\right)$ we will prove that the discrete Jensen's inequality holds at the connecting point of $f\big(T-1, \mathbf{s}'_k, d-1, L + \hat{L}\big)$ and $f\left(T-1, \mathbf{s}'_{d \beta_{dk} k}, k+1, L + \hat{L} - \beta_{dk}\right)$, i.e., at $\bar{L}=\beta_{dk}$. To be more specific, we will prove the following:
	\begin{align}\label{ineq:target_convex}
	f\left(T-1, \mathbf{s}'_k, d-1, \beta_{dk}\right) + f\left(T-1, \mathbf{s}'_{d \beta_{dk} k}, k+1, 2\right) \ge 2f\left(T-1, \mathbf{s}'_{d \beta_{dk} k}, k+1, 1\right).
	\end{align}
	From the definition of function $f$ in Eq. (\ref{F_def_remind}), we have
	\begin{align}
	f\left(T-1, \mathbf{s}'_k, d-1, \beta_{dk}\right) = J_{T-1}\left( \mathbf{s}'_{d \beta_{dk} k}\right),
	\end{align}
	and also
	\begin{align}
	f\left(T-1, \mathbf{s}'_{d \beta_{dk} k}, k+1, 0\right) = J_{T-1}\left( \mathbf{s}'_{d \beta_{dk} k}\right).
	\end{align}
	Hence,
	\begin{align}\label{eq:f_link}
	f\left(T-1, \mathbf{s}'_k, d-1, \beta_{dk}\right) = f\left(T-1, \mathbf{s}'_{d \beta_{dk} k}, k+1, 0\right).
	\end{align}
	Moreover, since $f\left(T-1, \mathbf{s}'_{d \beta_{dk} k}, k+1, L + \hat{L} \right) $ is discrete convex with respect to $L$ due to our assumption, we have
	\begin{align}\label{ineq:link_hold}
	f\left(T-1, \mathbf{s}'_{d \beta_{dk} k}, k+1, 0\right)  + f\left(T-1, \mathbf{s}'_{d \beta_{dk} k}, k+1, 2\right) \ge 2f\left(T-1, \mathbf{s}'_{d \beta_{dk} k}, k+1, 1\right).
	\end{align}
	By combining Eq. (\ref{eq:f_link}) and Ineq. (\ref{ineq:link_hold}), we prove that Ineq. (\ref{ineq:target_convex}) is true. Therefore, $g'_{dk}\left(L + \hat{L}\right)$ in Eq. (\ref{eq:Jf3}) is discrete convex with respect to $L$. Similarly, $g''_{dk}\left(L + \hat{L}\right)$ in Eq. (\ref{eq:Jf4}), $g'_{1k}\left(L + \hat{L}\right)$ in Eq. (\ref{eq:Jf5}), and $g''_{1k}\left(L + \hat{L}\right)$ in Eq. (\ref{eq:Jf6}) are also discrete convex with respect to $L$. 
	
	By using Eqs. (\ref{eq:Jf1})-(\ref{eq:Jf6}), we can express $G_{T-1}\left(\mathbf{s}, d, L + \hat{L}\right)$ defined in Eq. (\ref{G_new}) (where $\bar{L}$ is replaced by $L + \hat{L}$) by the sum of multiple functions $f$ having $T-1$ as the first argument. Furthermore, these functions are discrete convex with respect to $L$ according to our assumption at the beginning of this \textbf{Inductive step}. Therefore,  $G_{T-1}\left(\mathbf{s}, d, L + \hat{L}\right)$ is a discrete convex function with respect to $L$.
		
	Now, we will refer to the constant term $\mathds{1}_{d \ge 2}n_1C_p$ in (\ref{F_asmin}) as $constant$, and denote
	\begin{align}
		g\left(L + \hat{L}\right) = \left(L+\hat{L}\right)C_o + G_T\left(\mathbf{s}, d, L + \hat{L}\right).
	\end{align}
	where $G_T\left(\mathbf{s}, d, L + \hat{L}\right)$ has been proved to be a discrete convex function. In addition, $\left(L+\hat{L}\right)C_o$ is discrete linear, hence, also discrete convex. As a result, $g\left(L + \hat{L}\right)$ is a discrete convex function with respect to $L$. Then, Eq. (\ref{F_asmin}) becomes
	\begin{align}
		\underset{L \in \mathbb{L}_d\left(\mathbf{s}\right)}{f^{\text{A}}\left(T, \mathbf{s}, d, L \right)} = \underset{\hat{L} \in \left\{0, 1, \ldots, M_d-L\right\}}\min\left\{g\left(L+\hat{L} \right) \right\} + constant.
	\end{align}
	Assuming that $g\left(L+\hat{L} \right)$ attains its minimum $g^{\min}\left(L^*\right)$ at $L^*\ge n_1$. Note that the condition $L^* \ge n_1$ holds because all tasks having deadline 1 are excessive tasks, and will result in a penalty $C_p$ per task if not offloaded. Therefore, given that $C_p > C_o$, the optimal offloading decision cannot be less than $n_1$. Then, we have
	\begin{align}
		\underset{L \in \mathbb{L}_d\left(\mathbf{s}\right)}{f^{\text{A}}\left(T, \mathbf{s}, d, L \right)} =
		\begin{cases}
			g^{\min}\left(L^*\right) + constant, & \text{ if } L \le L^*, \\
			g\left(L\right) + constant \text{ where } \hat{L} = 0, & \text{ if } L > L^*.
		\end{cases}
	\end{align}
	As a results, $f^{\text{A}}\left(T, \mathbf{s}, d, L \right)$ are discrete convex functions with respect to $L \in \mathbb{L}_d\left(\mathbf{s}\right)$ for all states $\mathbf{s}$ and deadlines $d$. 
	
	$G_{T-1}\left(\mathbf{s}, d, L\right)$ in Eq. (\ref{F_noAMA}) can be expressed in a similar fashion as Eq. (\ref{G_new}) with $\bar{L}$ is replaced by $L$, since without the AMA, $\hat{L}=0$. Thus, $G_{T-1}\left(\mathbf{s}, d, L\right)$ in Eq. (\ref{F_noAMA}) can also be written as the sum of several functions $f$, each has $T-1$ as its first argument. This suggests that in Eq. (\ref{F_noAMA}), $G_{T-1}\left(\mathbf{s}, d, L\right)$ is a discrete convex function with respect to $L$ for every given $\mathbf{s}$ and $d$, and so is $f^{\overline{\text{A}}}\left(T, \mathbf{s}, d, L \right)$. Afterwards, by combining our definitions presented in Eqs. (\ref{FbarU_remind})-(\ref{FU_remind})  with the following equation:
	\begin{align}
		J_T\left(\bar{\mathbf{s}}_{dL}\right) = p_uJ_T^{\text{A}}\left(\bar{\mathbf{s}}_{dL}\right) + \left(1-p_u\right)J_T^{\overline{\text{A}}}\left(\bar{\mathbf{s}}_{dL}\right),
	\end{align}
	 we have
	\begin{align}\label{F_average_AMA}
		f\left(T, \mathbf{s}, d, L \right) = p_uf^{\text{A}}\left(T, \mathbf{s}, d, L \right) + \left(1-p_u\right)f^{\overline{\text{A}}}\left(T, \mathbf{s}, d, L \right).
	\end{align}
	Therefore, $f\left(T, \mathbf{s}, d, L \right)$ is also a discrete convex function with respect to $L$ for every given state $\mathbf{s}$ and deadline $d$.
	
	This result together with the presented \textbf{Initial case} lead us to a conclusion that $f\left(T, \mathbf{s}, d, L \right)$ are discrete convex functions with respect to $L$ for all parameters $T$, $\mathbf{s}$, and $d$. As $G_T\left(\mathbf{s}, d, L\right)$ is expressed by the sum of multiple functions $f$, so is $f^{\overline{\text{A}}}\left(T, \mathbf{s}, d, L \right)$, which is shown in Eq. (\ref{F_noAMA}). Therefore, $f^{\overline{\text{A}}}\left(T, \mathbf{s}, d, L \right)$ is also discrete convex functions with respect to $L$. Then, from our discussion at the beginning of this subsection, we can conclude that $F^{\overline{\text{A}}}\left(T, \mathbf{s}, d, L \right)$ are discrete convex functions with respect to $L$ for all parameters $T$, $\mathbf{s}$, and $d$. This completes our proof.

	\subsection{Proof of Lemma \ref{Lem:Convex2L*}}\label{proof_Lem:Convex2L*}
	We assume that function $F^{\overline{\text{A}}}\left(T, \mathbf{s}, d, L\right)$ attains its minimum at $L^*$ for $d=1$, the following set of inequalities hold
	\begin{align}\label{optim_fromF}
		J_T^{\overline{\text{A}}}\left(\bar{\mathbf{s}}_{1L^*}\right) + L^*C_o \le J_T^{\overline{\text{A}}}\left(\bar{\mathbf{s}}_{1L}\right) + LC_o, \text{ for all } L \in \mathbb{L}_1\left(\mathbf{s}\right).
	\end{align}
	
	From Eq. (\ref{JsL_intro}) and Ineqs. (\ref{optim_fromF}), we have
	\begin{align}
		J_T^{\text{A}}\left(\mathbf{s}, L^*\right) \le J_T^{\text{A}}\left(\mathbf{s}, L\right), \text{ for all } L \in \mathbb{L}_1\left(\mathbf{s}\right).
	\end{align}
	Equivalently,
	\begin{align}
		J_T^{\text{A}}\left(\mathbf{s}, L^*\right) = \underset{L \in \mathbb{L}_1\left(\mathbf{s}\right)}{\min} \left\{J_T^{\text{A}}\left(\mathbf{s}, L\right)\right\}.
	\end{align}
	We recall a fact that the optimal offloading decision must not be less than the number of tasks having deadline 1 which are all excessive tasks. Thus, $L^*$ is the optimal offloading decision of $\mathbf{s}$ for the time horizon $T$.
 
	\section{Appendix C: Proofs of Propositions}\label{Sec:AppendixPropositions}
	\subsection{Proof of Proposition \ref{Pro:SemiRedState}} \label{proof_Pro:SemiRedState}
	Considering a state $\mathbf{s} = \left(n_1, \ldots, n_N\right)$ and a time horizon $T$. Let $\mathbf{s}_r = \left(n^r_1, \ldots, n^r_N\right)$ be the corresponding reduced state obtained via Algorithm \ref{Gen2Red_Algo}, and let $\mathbf{s}_m = \left(n^m_1, \ldots, n^m_N\right)$ denote the corresponding lean state obtained according to Definition \ref{Def:SemRedStates}. As $n_i^m$, $i = 1, \ldots, N$, are defined by Eq. (\ref{eq:semi_element_def}) with the parameters $\gamma_i$, $i = 1, \ldots, N$, are given in Eq. (\ref{eq:gamma_def}), we have
\begin{align}
n_i^r \le n_i^m \le n_i, ~ i = 1, \ldots, N.
\end{align}
Moreover, we recall that $n_i - n^r_i$ tasks having deadline $i$ in $\mathbf{s}$, $ i = 1, \ldots, N$ are excessive tasks, i.e., they are guaranteed to expire if not offloaded within the first $i$ time slots. Therefore, $n_i - n^m_i$ tasks having deadline $i$ in $\mathbf{s}$, $ i = 1, \ldots, N$ are also excessive tasks. In other words, for each deadline, tasks that $\mathbf{s}$ has more than $\mathbf{s}_m$ are excessive tasks.
\begin{example}
    For example, with $\mathbf{s} = \left(0, 3, 4, 0, 5 \right)$, the corresponding lean state would be $\mathbf{s}_r= \left(0, 1, 1, 0, 4 \right)$. Then, for deadline 1, there are $3-1=2$ excessive tasks. For deadline 2, there are $4-1=3$ excessive tasks. For deadline 4 and 5, there is $0$ excessive task.
\end{example}
	Now let us consider the following cases in which we use the notations
	$J_T\left(\mathbf{s}\right) \big|_{j}$ and $J_T\left(\mathbf{s}_m\right) \big|_{j}$ to
	denote the minimum average cost of $\mathbf{s}$ and $\mathbf{s}_m$, respectively, given that the AMA arrives
	for the first time at time slot $t$. Then, there are following cases.
	\begin{itemize}
		\item Case 0: The AMA is available for the first time at the current time slot, $t=0$. As we mentioned, for every deadline, tasks that state $\mathbf{s}$ has more than state $\mathbf{s}_m$ are excessive tasks which should be offloaded by the optimal policy whenever the AMA is available. Therefore, in this case, if $L^*_m$ denotes the optimal number of tasks to offload of $\mathbf{s}_m$, that of $\mathbf{s}$ will be $L^*_m + y$, where $y = \sum_{i=1}^N (n_i-n_i^m)$ is the partial number of excessive tasks in $\mathbf{s}$. Hence,
		\begin{align}
			J_T\left(\mathbf{s}\right)\big|_{0} = J_T\left(\mathbf{s}_m\right)\big|_{0} + C_o\sum_{i=1}^N\left(n_i-n_i^m\right).
		\end{align}
		
		\item Case 1: If the AMA is available for the first time at time slot 1, the number of tasks having deadline 1 expiring from $\mathbf{s}$ is more than that from $\mathbf{s}_m$ by $n_{1} - n_{1}^m$. All the other remaining excessive
		tasks can be offloaded from both states $\mathbf{s}$ and $\mathbf{s}_m$. Hence,
		\begin{align}
			J_T\left(\mathbf{s}\right)\big|_{1} = J_T\left(\mathbf{s}_m\right)\big|_{1} + C_o\sum_{i=2}^N\left(n_i-n_i^m\right) + \left(n_{1} - n_{1}^m\right)C_p.
		\end{align}
		
		The same logic can be applied to other cases when the AMA
		first arrives at time slot $2, 3, \ldots, N$. Therefore, we present next the last case.
		
		\item Case $N$:  If the AMA is available for the
		first time at time slot $N$, the number of tasks having deadline $i$ expiring from $\mathbf{s}$ is more than that from $\mathbf{s}_m$ by $n_i - n_i^m$. Therefore, we have
		\begin{align}
			J_T\left(\mathbf{s}\right)\big|_{N} = J_T\left(\mathbf{s}_m\right)\big|_{N} + C_p\sum_{i=1}^N\left(n_i-n_i^m\right).
		\end{align}
	\end{itemize}
	
	The probability that the AMA arrives for the first time at time slot $t$ is computed as
	\begin{align}
		P_t = p_u\left(1-p_u \right)^{t}.
	\end{align}
	Also, the probability that the AMA does not arrive within the first $N$ time slots is
	\begin{align}
		P_{t \ge N} = \left(1-p_u \right)^{N}.
	\end{align}
	From the above logic, $J_T\left(\mathbf{s} \right)$ can be expressed in terms of $J_T\left(\mathbf{s}_m\right)$ as follows:
	\begin{align} \label{proof_expand}
		\begin{split}
			J_T\left(\mathbf{s}\right) &=  P_{t=0}\left(J_T\left(\mathbf{s}_m\right)\big|_{0} + C_o\sum_{i=1}^{N} \left(n_i - n_i^m \right) \right)\\
			& + P_{t=1}\left( J_T\left(\mathbf{s}_m\right)\big|_{1}  + C_o\sum_{i=2}^{N} \left(n_i - n_i^m \right) + C_p\left(n_1 - n_1^m \right)\right)\\
			& + \ldots \\
			& + P_{t= N-1}\left( J_T\left(\mathbf{s}_m\right)\big|_{N-1}  + C_o \left(n_N - n_N^m \right) + C_p\sum_{i=1}^{N-1} \left(n_i - n_i^m \right)\right) \\
			& + P_{t \ge N} \left(J_T\left(\mathbf{s}_m\right)\big|_{t\geq N}  + C_p\sum_{i=1}^{N} \left(n_i - n_i^m \right) \right).
		\end{split}
	\end{align}
	$J_T\left(\mathbf{s}_m\right)$ is the minimum cost averaged over
	all cases, hence, can be expressed by
	\begin{align} \label{proof_expla}
		J_T\left(\mathbf{s}_m\right) = \sum_{i=0}^{N-1} P_{t=i} J_T\left(\mathbf{s}_m\right) \big|_{i} + P_{t \ge N} J_T\left(\mathbf{s}_m\right) \big|_{t\geq N}.
	\end{align}
	In conclusion, with the aid of Eq. (\ref{proof_expla}), the equation (\ref{proof_expand}) can be simplified to Eq. (\ref{eq:gen2sem_eq}).

	\subsection{Proof of Proposition \ref{Pro:deadline_postponed}} \label{proof_Pro:deadline_postponed}
	Given an initial state $\mathbf{s} = \left(n_1, \ldots, n_N\right) \not\equiv \left(0, \ldots, 0\right)$, and a time horizon $T$. We call $d$ the deadline of the most imminent task of $\mathbf{s}$, i.e., $d$ is the smallest deadline such that $n_d > 0$. Denoting $\tilde{\mathbf{s}} = \left(\tilde{n}_1, \ldots, \tilde{n}_N \right) \in \mathbb{S}_{pp}\left(\mathbf{s}\right)$. Trivially, we have
	\begin{align}
		J_T\left(\tilde{\mathbf{s}}\right) = J_T\left(\mathbf{s}\right), \text{ if } N=1.
	\end{align}
	
	If $N \ge 2$, by definition of set $\mathbb{S}_{pp}\left(\mathbf{s}\right)$, there must be a deadline $\tilde{d} \in \left\{d, \ldots, N\right\}$ such that
	\begin{align}\label{element_stil}
		\tilde{n}_i = 
		\begin{cases}
			n_i, &\text{ if } i \ne \tilde{d} \text{ and } i \ne d,\\
			n_d - 1, &\text{ if } i = d,\\
			n_{\tilde{d}} + 1, &\text{ if } i = \tilde{d}.
		\end{cases}
	\end{align}
	
	If $\tilde{d} = d$, $\mathbf{s} \equiv \tilde{\mathbf{s}}$. It is trivially that
	\begin{align}
		J_T\left(\tilde{\mathbf{s}}\right) = J_T\left(\mathbf{s}\right), \text{ if } N \ge 2 \text{ and } \tilde{d} = d.
	\end{align}

	For the rest of this part, we will prove that
	\begin{align}
	J_T\left(\tilde{\mathbf{s}}\right) \le J_T\left(\mathbf{s}\right), \text{ when } \tilde{d} \ge d+1, d \le N-1, N \ge 2.
	\end{align}
	We conduct our proof via induction starting with an initial case.
	
	\textbf{Initial case:} Given $\mathbf{s} = \left(n_1, \ldots, n_N\right)$, $\tilde{\mathbf{s}} = \left(\tilde{n}_1, \ldots, \tilde{n}_N \right) \in \mathbb{S}_{pp}\left(\mathbf{s}\right)$, and $T=1$.
	\begin{itemize}
		\item If $n_1 = 0$, $\tilde{n}_1 = 0$ according to Eq. (\ref{element_stil}). Therefore,
		\begin{align}
			J_{1}\left(\tilde{\mathbf{s}}\right) = J_{1}\left(\mathbf{s}\right) = 0.
		\end{align}
	
		\item If $n_1 \ge 1$, $\tilde{n}_1 = n_1-1$ according to Eq. (\ref{element_stil}). We have
		\begin{align}
			J_{1}\left(\mathbf{s}\right) & = \left(p_uC_o + \left(1-p_u\right)C_p\right)n_1,\\
			J_{1}\left(\tilde{\mathbf{s}}\right) &=\left(p_uC_o + \left(1-p_u\right)C_p\right)\left(n_1-1\right).
		\end{align}
		Therefore,
		\begin{align}
			J_{1}\left(\tilde{\mathbf{s}}\right) \le J_{1}\left(\mathbf{s}\right).
		\end{align}
	\end{itemize}
	
	The subsequent part is an inductive step.
	
	\textbf{Inductive step:} In this step, we make the following assumption: \textit{For every given state $\mathbf{s}$ and $\tilde{\mathbf{s}} \in \mathbb{S}_{pp}\left(\mathbf{s}\right)$, the following holds}
	\begin{align}\label{ineq:assume_pp}
	J_{T-1}\left(\tilde{\mathbf{s}}\right) \le J_{T-1}\left(\mathbf{s}\right),
	\end{align}
	
	Initially, we highlight an observation in the following remark. 
	\begin{remark}\label{remark_pp}
		\textit{Given two states $\mathbf{s}$ and $\tilde{\mathbf{s}}$ which are assumed to transit to $\mathbf{s}'$ and $\tilde{\mathbf{s}}'$ in the next time slot, respectively, if no offloading is performed. If $\tilde{\mathbf{s}} \in \mathbb{S}_{pp}\left(\mathbf{s}\right)$ and the number of tasks having deadline 1 of $\mathbf{s}$ is 0, then, $\tilde{\mathbf{s}}' \in \mathbb{S}_{pp}\left(\mathbf{s}'\right)$.}
	\end{remark}
	Remark \ref{remark_pp} can be proved as follows. Given state $\mathbf{s}$ and $\tilde{\mathbf{s}} \in \mathbb{S}_{pp}\left(\mathbf{s}\right)$ as above with $n_1 = 0$. We assume that no offloading is performed, and that $\mathbf{s}$ and $\tilde{\mathbf{s}}$ becomes $\mathbf{s}_{ds} = \left(n^{ds}_1, \ldots, n^{ds}_N\right)$ and $\tilde{\mathbf{s}}_{ds} = \left(\tilde{n}^{ds}_2, \ldots, \tilde{n}^{ds}_N\right)$, respectively, after the deadline shifting, in which
	\begin{align}\label{cond:pp_match1}
		n^{ds}_i = 
		\begin{cases}
			n_{i+1}, &\text{ if } i \ne N,\\
			0, &\text{ if } i = N,
		\end{cases}
	\end{align}
	and
	\begin{align}\label{cond:pp_match2}
		\tilde{n}^{ds}_i = 
		\begin{cases}
			n^{ds}_{i}, &\text{ if } i \ne \tilde{d}-1 \text{ and } i \ne d-1,\\
			n^{ds}_d - 1, &\text{ if } i = d-1,\\
			n^{ds}_{\tilde{d}} + 1, &\text{ if } i = \tilde{d}-1,\\
			0, &\text{ if } i = N.
		\end{cases}
	\end{align}
	Eqs. (\ref{cond:pp_match1}) and (\ref{cond:pp_match2}) indicate that $\tilde{\mathbf{s}}_{ds}\in \mathbb{S}_{pp}\left(\mathbf{s}_{ds}\right)$. Therefore, with the same realizations of task arrival and local processing, $\mathbf{s}_{ds}$ and $\tilde{\mathbf{s}}_{ds}$ would transit to states $\mathbf{s}'$ and $\tilde{\mathbf{s}}'$, respectively, where $\tilde{\mathbf{s}}' \in \mathbb{S}_{pp}\left(\mathbf{s}'\right)$.
	
	Based on the emphasized observation above, let us consider the following two cases:\\
	\textbf{Case 1}: $\mathbf{s}$ is a non-offloading state.\\
	\textbf{Case 2}: The AMA does not presents at the initial time slot, and $n_1=0$.

	We note that in \textbf{Case 1}, $\mathbf{s}$ is a non-offloading state, implying that $n_1=0$. This is because tasks having deadline 1 are all excessive tasks and must be offloaded whenever the AMA is available. Applying the DP equation (\ref{DP_Eq}) in Case 1, we have
	\begin{align}
		J_T\left(\mathbf{s}\right) & =  \mu \sum_{k=0}^N p_k J_{T-1}\left(\mathbf{s}'_{0k} \right) + \left(1-\mu\right)\sum_{k=0}^N p_k J_{T-1}\left(\mathbf{s}''_{0k}\right), \label{eq:Js_case1}\\
		J_T\left(\tilde{\mathbf{s}}\right) &=  \mu \sum_{k=0}^N p_k J_{T-1}\left(\tilde{\mathbf{s}}'_{0k} \right) + \left(1-\mu\right)\sum_{k=0}^N p_k J_{T-1}\left(\tilde{\mathbf{s}}''_{0k}\right). \label{eq:Jstil_case1}
	\end{align}
	Applying Eq. (\ref{eq:JNoAMA}) to \textbf{Case 2} yields
	\begin{align}
		J_T^{\overline{\text{A}}}\left(\mathbf{s}\right) & =  \mu \sum_{k=0}^N p_k J_{T-1}\left(\mathbf{s}'_{0k} \right) + \left(1-\mu\right)\sum_{k=0}^N p_k J_{T-1}\left(\mathbf{s}''_{0k}\right), \label{eq:Js_case2}\\
		J_T^{\overline{\text{A}}}\left(\tilde{\mathbf{s}}\right) & =  \mu \sum_{k=0}^N p_k J_{T-1}\left(\tilde{\mathbf{s}}'_{0k} \right) + \left(1-\mu\right)\sum_{k=0}^N p_k J_{T-1}\left(\tilde{\mathbf{s}}''_{0k}\right). \label{eq:Jstil_case2}.
	\end{align} 
	As both \textbf{Case 1} and \textbf{Case 2} match the condition mentioned in Remark \ref{remark_pp} that $\tilde{\mathbf{s}} \in \mathbb{S}_{pp}\left(\mathbf{s}\right)$ and $n_1=0$, we have: $\tilde{\mathbf{s}}'_{0k} \in \mathbb{S}_{pp}\left(\mathbf{s}'_{0k}\right)$ and $\tilde{\mathbf{s}}''_{0k} \in \mathbb{S}_{pp}\left(\mathbf{s}''_{0k}\right)$. Combining this with our assumption in Eq. (\ref{ineq:assume_pp}), we have
	\begin{align}
	J_{T-1}\left( \tilde{\mathbf{s}}'_{0k}\right) \le J_{T-1}\left( \mathbf{s}'_{0k}\right) \label{ineq:recursive_pp1}, \\
	J_{T-1}\left( \tilde{\mathbf{s}}''_{0k}\right) \le J_{T-1}\left( \mathbf{s}''_{0k}\right). \label{ineq:recursive_pp2}
	\end{align}
	
	Combining Ineqs. (\ref{ineq:recursive_pp1})-(\ref{ineq:recursive_pp2}) with Eqs. (\ref{eq:Js_case1})-(\ref{eq:Jstil_case1}) gives
	\begin{align}\label{ineq:proved_nonoff}
	J_T\left(\tilde{\mathbf{s}}\right) \le J_T\left(\mathbf{s}\right), \text{ when } \mathbf{s} \text{ is a non-offloading state.}
	\end{align}
	Combining Ineqs. (\ref{ineq:recursive_pp1})-(\ref{ineq:recursive_pp2}) with Eqs. (\ref{eq:Js_case2})-(\ref{eq:Jstil_case2}) gives
	\begin{align}\label{ineq:proved_NoU_0}
	J_T^{\overline{\text{A}}}\left(\tilde{\mathbf{s}}\right) \le J_T^{\overline{\text{A}}}\left(\mathbf{s}\right), \text{ when } n_1=0.
	\end{align}
	
	The other case without AMA's presence is as follows:\\
	\textbf{Case 3}: The AMA does not presents at the initial time slot, and $n_1 \ge 1$.
	
	In this case, assuming that $\mathbf{s}$ and $\tilde{\mathbf{s}}$ transit to $\mathbf{s}' = \left(n'_1, \ldots, n'_N\right)$ and $\tilde{\mathbf{s}}' = \left(\tilde{n}'_1, \ldots, \tilde{n}'_N\right)$ in the next time slot, respectively. Based on Eq. (\ref{element_stil}), we have
	\begin{align}\label{element_stil1}
		\tilde{n}'_i = 
		\begin{cases}
			n'_{i}, &\text{ if } i \ne \tilde{d}-1,\\
			n'_{\tilde{d}} + 1, &\text{ if } i = \tilde{d}-1.
		\end{cases}
	\end{align}
	We note that all tasks having deadline 1 cannot be processed and will expire, since the deadline shifting happens before the local processing event as presented in Fig.~\ref{fig:order-events}. Therefore, from Eq. (\ref{element_stil}), there is a fact that after the first time slot, the number of tasks expiring from $\mathbf{s}$ is more than that from $\tilde{\mathbf{s}}$ by 1. This observation leads to the following equation.
	\begin{align}\label{eq:ss_nextss}
		J_T^{\overline{\text{A}}}\left(\mathbf{s}\right) - J_T^{\overline{\text{A}}}\left(\tilde{\mathbf{s}}\right) = C_p + J_T^{\overline{\text{A}}}\left(\mathbf{s}'\right) - J_T^{\overline{\text{A}}}\left(\tilde{\mathbf{s}}'\right).
	\end{align}
	From Eq. (\ref{element_stil1}), we can also observe that $\tilde{\mathbf{s}}'$ only differs from $\mathbf{s}'$ by an additional task having deadline $\tilde{d}$. Furthermore, $C_p$ is the largest cost that can possibly be incurred by a task. We have the following inequality.
	\begin{align}\label{ineq:ss_nextss}
		C_p + J_T^{\overline{\text{A}}}\left(\mathbf{s}'\right) - J_T^{\overline{\text{A}}}\left(\tilde{\mathbf{s}}'\right) \ge 0.
	\end{align}
	Applying Ineq. (\ref{ineq:ss_nextss}) to Eq. (\ref{eq:ss_nextss}) gives us
	\begin{align}\label{ineq:proved_NoU_1}
		J_T^{\overline{\text{A}}}\left(\tilde{\mathbf{s}}\right) \le J_T^{\overline{\text{A}}}\left(\mathbf{s}\right) \text{ when } n_1 \ge 1.
	\end{align}
	From Ineqs. (\ref{ineq:proved_NoU_0}) and (\ref{ineq:proved_NoU_1}), we have
	\begin{align}\label{ineq:proved_NoU}
		J_T^{\overline{\text{A}}}\left(\tilde{\mathbf{s}}\right) \le J_T^{\overline{\text{A}}}\left(\mathbf{s}\right).
	\end{align}
	
	Finally, it remains to consider the following case.\\
	\textbf{Case 4}: The AMA presents at the initial time slot, and $\mathbf{s}$ is an offloading state.
	
	In this case, we consider two contexts in which both are associated with a time horizon $T$. In the first context, state $\mathbf{s}$ is given initially, and $L^*$ is the corresponding optimal offloading decision. In the second one, $\tilde{\mathbf{s}}$ is given as an initial state, and task offloading is done according to a policy $\pi\left(L^*\right)$ in the first time slot. The task offloading decision of policy $\pi\left(L^*\right)$ in the first time slot is defined as follows.
	\begin{itemize}
		\item Situation 1: $L^* \ge 1 + \sum_{i=1}^{\tilde{d}-1} n_i$, then, $\pi\left(L^*\right)$ offloads $L^*$ most imminent tasks from $\tilde{\mathbf{s}}$.
		\item Situation 2: $0 < L^* < 1 + \sum_{i=1}^{\tilde{d}-1} n_i$, then, $\pi\left(L^*\right)$ offloads $L^*-1$ most imminent tasks, and offloads a task with deadline $\tilde{d}$ from $\tilde{\mathbf{s}}$.
	\end{itemize}
	For the two situations mentioned above, the policy $\pi\left(L^*\right)$ guarantees that in both contexts, we arrive at the same state after the offloading step. For intuition, we consider the following example which corresponds to Situation 1.
	\begin{example}
		Given $\mathbf{s} = \left(0, 0, 5, 4, 5\right)$ and $\tilde{\mathbf{s}} = \left(0, 0, 4, 4, 6\right) \in \mathbb{S}_{pp}\left(\mathbf{s}\right)$ where $\tilde{d} = 5$. Assuming that 12 most imminent tasks are removed from $\mathbf{s}$ resulting in state $\left(0, 0, 0, 0, 2\right)$. Since $12 > 1 + 5 + 4$, we remove 12 most imminent tasks from $\tilde{\mathbf{s}}$, which also results in state $\left(0, 0, 0, 0, 2\right)$.
	\end{example}
	The next example corresponding to Situation 2 is as follows.
	\begin{example}
		Given $\mathbf{s} = \left(0, 0, 5, 1, 1\right)$ and $\tilde{\mathbf{s}} = \left(0, 0, 4, 1, 2\right) \in \mathbb{S}_{pp}\left(\mathbf{s}\right)$ where $\tilde{d} = 5$. Assuming that 5 most imminent tasks are removed from $\mathbf{s}$ resulting in state $\left(0, 0, 0, 1, 1\right)$. Since $0 < 5 < 1 + 5 + 1$, we remove $5-1=4$ most imminent tasks, and a task with deadline $\tilde{d} = 5$ from $\tilde{\mathbf{s}}$, which also gives state $\left(0, 0, 0, 1, 1\right)$.
	\end{example}
	
	With the AMA's availability, we denote $J_{T, \pi}^{\text{A}}\left(\tilde{\mathbf{s}}\right)$ the minimum average cost associated with state $\tilde{\mathbf{s}}$ and time horizon $T$ given that the task offloading is done according to policy $\pi\left(L^*\right)$ at the initial time slot. Since in both contexts described above, we arrive at the same state after the task offloading step by offloading the same number of tasks, we have the following equality
	\begin{align}
		J_{T,\pi}^{\text{A}}\left(\tilde{\mathbf{s}}\right) = J_T^{\text{A}}\left(\mathbf{s}\right).
	\end{align}
	Trivially, $J_T^{\text{A}}\left(\tilde{\mathbf{s}}\right) \le J_{T, \pi}^{\text{A}}\left(\tilde{\mathbf{s}}\right)$. Thus, we have
	\begin{align}\label{ineq:proved_U}
		J_T^{\text{A}}\left(\tilde{\mathbf{s}}\right) \le  J_T^{\text{A}}\left(\mathbf{s}\right).
	\end{align}
	
	We recall the following expression:
	\begin{align}\label{eq:J_Jcond}
	J_T\left(\mathbf{s}\right) = p_uJ_T^{\text{A}}\left(\mathbf{s}\right) + \left(1-p_u\right)J_T^{\overline{\text{A}}}\left(\mathbf{s}\right) \text{ for every state } \mathbf{s}.
	\end{align}
	Combining Ineqs. (\ref{ineq:proved_NoU}) and (\ref{ineq:proved_U}) with (\ref{eq:J_Jcond}) gives us
	\begin{align}\label{ineq:proved_off}
	J_T\left(\tilde{\mathbf{s}}\right) \le J_T\left(\mathbf{s}\right), \text{ when } \mathbf{s} \text{ is an offloading state.}
	\end{align}
	
	Ineqs. (\ref{ineq:proved_nonoff}) and (\ref{ineq:proved_off}) lead us to the following inequality
	\begin{align}\label{ineq:end_induction}
	J_T\left(\tilde{\mathbf{s}}\right) \le J_T\left(\mathbf{s}\right).
	\end{align}
	
	Combining Ineq. (\ref{ineq:end_induction}) with the presented \textbf{Initial case} allows us to conclude that
	\begin{align}
	J_T\left(\tilde{\mathbf{s}}\right) \le J_T\left(\mathbf{s}\right), \text{ for every state } \mathbf{s} \text{ and } \tilde{\mathbf{s}} \in \mathbb{S}_{pp}\left(\mathbf{s}\right).
	\end{align}
		
	\subsection{Proof of Proposition \ref{Pro:ONO_conds}}\label{proof_Pro:ONO_conds}
	Given a time horizon $T$, if $\mathbf{s}_a$ is a non-offloading state, the corresponding optimal offloading decision is 0, thus,
	\begin{align}\label{ineq_NO1}
		J_T^{\text{A}}\left(\mathbf{s}_a, 0\right) < J_T^{\text{A}}\left(\mathbf{s}_a, 1\right).
	\end{align}
	From Eq. (\ref{JsL_intro}), the above inequality is equivalent to
	\begin{align}\label{ineq_NO}
		J_T^{\overline{\text{A}}}\left(\mathbf{s}_a\right) < J_T^{\overline{\text{A}}}\left(\mathbf{s}\right) + C_o.
	\end{align}
	Here, we have $\mathbf{s}_a \in \mathbb{S}_{adj}\left(\mathbf{s}\right)$. According to Theorem \ref{Theo:adjacent_L*}, if $\mathbf{s}_a$ is a non-offloading state, $\mathbf{s}$ is also a non-offloading state. Hence, trivially, we have
	\begin{align}
		J_T^{\text{A}}\left(\mathbf{s}_a\right) &= J_T^{\overline{\text{A}}}\left(\mathbf{s}_a\right), \label{eq:U=Ubar1} \\
		J_T^{\text{A}}\left(\mathbf{s}\right) &= J_T^{\overline{\text{A}}}\left(\mathbf{s}\right). \label{eq:U=Ubar2}
	\end{align}
	Combining Eqs. (\ref{eq:U=Ubar1}) and (\ref{eq:U=Ubar2}) with Eq. (\ref{gen_U_noU}) and Ineq. (\ref{ineq_NO}), we have
	\begin{align}\label{optimcond_NO}
		J_T\left(\mathbf{s}_a\right) - J_T\left(\mathbf{s}\right) < C_o.
	\end{align}
	We make a conclusion at this point that: For two states $\mathbf{s}$ and $\mathbf{s}_a \in \mathbb{S}_{adj}\left(\mathbf{s}\right)$, if $\mathbf{s}_a$ is a non-offloading state, Ineq. (\ref{optimcond_NO}) holds.
	
	Now, we will prove the reverse in which there are two states $\mathbf{s}$ and $\mathbf{s}_a \in \mathbb{S}_{adj}\left(\mathbf{s}\right)$, and the inequality (\ref{optimcond_NO}) holds. We assume, in contradict, that $\mathbf{s}_a$ is an offloading state. By applying Eq. (\ref{gen_U_noU}) to Ineq. (\ref{optimcond_NO}), we have
	\begin{align}\label{ineq:NOconds_rev}
	p_u\left( J_T^{\text{A}}\left(\mathbf{s}_a\right) - J_T^{\text{A}}\left(\mathbf{s}\right) \right) + \left(1-p_u\right)\left( J_T^{\overline{\text{A}}}\left(\mathbf{s}_a\right) - J_T^{\overline{\text{A}}}\left(\mathbf{s}\right) \right) < C_o.
	\end{align}
	We denote $L_a^*  \ge 1$ the optimal offloading decision of $\mathbf{s}_a$. From Theorem \ref{Theo:adjacent_L*}, the optimal offloading decision of $\mathbf{s}$ would be $L_a^*-1$. Combining this with Eq. (\ref{rel_orig}), we have
	\begin{align}
	J_T^{\text{A}}\left(\mathbf{s}_a\right) &= J_T^{\text{A}}\left(\mathbf{s}_a, L_a^*\right) = J_T^{\text{A}}\left(\bar{\mathbf{s}}_{a1L_a^*}, 0\right) + L_a^*C_o, \\
	J_T^{\text{A}}\left(\mathbf{s}\right) &= J_T^{\text{A}}\left(\mathbf{s}, L_a^*-1\right) = J_T^{\text{A}}\left(\bar{\mathbf{s}}_{1\left(L_a^*-1\right)}, 0\right) + \left(L_a^*-1\right)C_o,
	\end{align}
	where state $\bar{\mathbf{s}}_{a1L_a^*}$ is obtained by offloading $L_a^*$ most imminent tasks from $\mathbf{s}_a$, and $\bar{\mathbf{s}}_{1\left(L_a^*-1\right)}$ is obtained by offloading $L_a^*-1$ most imminent tasks from $\mathbf{s}$. 
	
	We recall that
	\begin{align}
	 	J_T^{\text{A}}\left(\mathbf{s}, L\right) = \mathcal{C}^{\text{A}}\left(\mathbf{s}, L\right) + G_T^{\text{A}}\left(\mathbf{s}, L\right)
	 \end{align}
	  denotes the minimum average cost attained over $T$ time slots by offloading $L$ most imminent tasks from $\mathbf{s}$ given the AMA's availability. From the definition of adjacent states in Definition \ref{Def:adjacent_states}, we have that $\bar{\mathbf{s}}_{a1L_a^*} \equiv \bar{\mathbf{s}}_{1\left(L_a^*-1\right)}$. As a result,
	\begin{align}
	J_T^{\text{A}}\left(\mathbf{s}_a\right) - J_T^{\text{A}}\left(\mathbf{s}\right) = C_o.
	\end{align}
	
	Combining the above equality with Ineq. (\ref{ineq:NOconds_rev}), we have
	\begin{align}
	J_T^{\overline{\text{A}}}\left(\mathbf{s}_a\right) - J_T^{\overline{\text{A}}}\left(\mathbf{s}\right)  < C_o.
	\end{align}
	Using the fact that $\mathbf{s}$ is obtained by offloading the most imminent task from $\mathbf{s}_a$, the above inequality can be re-written as
	\begin{align}
	J_T^{\overline{\text{A}}}\left(\bar{\mathbf{s}}_{a10}\right)  < J_T^{\overline{\text{A}}}\left(\bar{\mathbf{s}}_{a11}\right) + C_o,
	\end{align}
	in which state $\bar{\mathbf{s}}_{a10}$ and $\bar{\mathbf{s}}_{a11}$ are obtained by offloading 0 task, and offloading the most imminent task from $\mathbf{s}_a$, respectively. From the definition of function $F^{\overline{\text{A}}}$ in Eq. (\ref{F_def}), the above inequality is equivalent to
	\begin{align}\label{ineq:ineq_trigger}
	F^{\overline{\text{A}}}\left(T, \mathbf{s}_a, 1, 0\right)  < F^{\overline{\text{A}}}\left(T, \mathbf{s}_a, 1, 1\right).
	\end{align}
	Now, combining Ineq. (\ref{ineq:ineq_trigger}) with the convexity of function $F^{\overline{\text{A}}}$ presented in Lemma \ref{Lem:Convexity}, we have
	\begin{align}
	F^{\overline{\text{A}}}\left(T, \mathbf{s}_a, 1, 0\right)  < F^{\overline{\text{A}}}\left(T, \mathbf{s}_a, 1, L\right) + LC_o, \text{ for all } L \in \mathbb{L}_1\left(\mathbf{s}_a\right).
	\end{align}
	This suggests that 0 is the optimal offloading decision associated with state $\mathbf{s}_a$ for the given time horizon $T$, which is contradict to our assumption that $\mathbf{s}_a$ is an offloading state. Therefore, we can make a conclusion that: As Ineq. (\ref{optimcond_NO}) holds, $\mathbf{s}_a$ is a non-offloading state.
	
	To this end, the following has been proved: For two states $\mathbf{s}$ and $\mathbf{s}_a \in \mathbb{S}_{adj}\left(\mathbf{s}\right)$, state $\mathbf{s}_a$ is a non-offloading state if and only if Ineq. (\ref{optimcond_NO}) holds. As a consequence, $\mathbf{s}_a$ is an offloading state if and only if $J_T\left(\mathbf{s}_a\right) - J_T\left(\mathbf{s}\right) \ge C_o$.

	\bibliographystyle{IEEEtran}
	\bibliography{IEEEabrv,DNK}

\begin{thebibliography}{10}
\providecommand{\url}[1]{#1}
\csname url@samestyle\endcsname
\providecommand{\newblock}{\relax}
\providecommand{\bibinfo}[2]{#2}
\providecommand{\BIBentrySTDinterwordspacing}{\spaceskip=0pt\relax}
\providecommand{\BIBentryALTinterwordstretchfactor}{4}
\providecommand{\BIBentryALTinterwordspacing}{\spaceskip=\fontdimen2\font plus
\BIBentryALTinterwordstretchfactor\fontdimen3\font minus
  \fontdimen4\font\relax}
\providecommand{\BIBforeignlanguage}[2]{{%
\expandafter\ifx\csname l@#1\endcsname\relax
\typeout{** WARNING: IEEEtran.bst: No hyphenation pattern has been}%
\typeout{** loaded for the language `#1'. Using the pattern for}%
\typeout{** the default language instead.}%
\else
\language=\csname l@#1\endcsname
\fi
#2}}
\providecommand{\BIBdecl}{\relax}
\BIBdecl

\bibitem{mao2017survey}
Y.~Mao, C.~You, J.~Zhang, K.~Huang, and K.~B. Letaief, ``A survey on mobile
  edge computing: The communication perspective,'' \emph{IEEE communications
  surveys \& tutorials}, vol.~19, no.~4, pp. 2322--2358, 2017.

\bibitem{dinh2013survey}
H.~T. Dinh, C.~Lee, D.~Niyato, and P.~Wang, ``A survey of mobile cloud
  computing: architecture, applications, and approaches,'' \emph{Wireless
  communications and mobile computing}, vol.~13, no.~18, pp. 1587--1611, 2013.

\bibitem{fragkos2020artificial}
G.~Fragkos, E.~E. Tsiropoulou, and S.~Papavassiliou, ``Artificial intelligence
  enabled distributed edge computing for internet of things applications,'' in
  \emph{2020 16th international conference on distributed computing in sensor
  systems (DCOSS)}.\hskip 1em plus 0.5em minus 0.4em\relax IEEE, May 2020, pp.
  450--457.

\bibitem{teymoori2020efficient}
P.~Teymoori, T.~D. Todd, D.~Zhao, and G.~Karakostas, ``Efficient mobile
  computation offloading with hard task deadlines and concurrent local
  execution,'' in \emph{GLOBECOM 2020-2020 IEEE Global Communications
  Conference}.\hskip 1em plus 0.5em minus 0.4em\relax IEEE, 2020, pp. 1--6.

\bibitem{geng2018energy}
Y.~Geng, Y.~Yang, and G.~Cao, ``Energy-efficient computation offloading for
  multicore-based mobile devices,'' in \emph{IEEE INFOCOM 2018-IEEE Conference
  on Computer Communications}.\hskip 1em plus 0.5em minus 0.4em\relax IEEE,
  2018, pp. 46--54.

\bibitem{zhang2014collaborative}
W.~Zhang, Y.~Wen, and D.~O. Wu, ``Collaborative task execution in mobile cloud
  computing under a stochastic wireless channel,'' \emph{IEEE Transactions on
  Wireless Communications}, vol.~14, no.~1, pp. 81--93, 2014.

\bibitem{huang2021deadline}
H.~Huang, Q.~Ye, and Y.~Zhou, ``Deadline-aware task offloading with
  partially-observable deep reinforcement learning for multi-access edge
  computing,'' \emph{IEEE Transactions on Network Science and Engineering},
  2021.

\bibitem{du2017computation}
J.~Du, L.~Zhao, J.~Feng, and X.~Chu, ``Computation offloading and resource
  allocation in mixed fog/cloud computing systems with min-max fairness
  guarantee,'' \emph{IEEE Transactions on Communications}, vol.~66, no.~4, pp.
  1594--1608, 2017.

\bibitem{van2017optimization}
D.~Van~Le and C.-K. Tham, ``An optimization-based approach to offloading in
  ad-hoc mobile clouds,'' in \emph{GLOBECOM 2017-2017 IEEE Global
  Communications Conference}.\hskip 1em plus 0.5em minus 0.4em\relax IEEE,
  2017, pp. 1--6.

\bibitem{van2018deep}
------, ``A deep reinforcement learning based offloading scheme in ad-hoc
  mobile clouds,'' in \emph{IEEE INFOCOM 2018-IEEE Conference on Computer
  Communications Workshops (INFOCOM WKSHPS)}.\hskip 1em plus 0.5em minus
  0.4em\relax IEEE, 2018, pp. 760--765.

\bibitem{stanley2015catalan}
R.~P. Stanley, \emph{Catalan numbers}.\hskip 1em plus 0.5em minus 0.4em\relax
  Cambridge University Press, 2015.

\end{thebibliography}

\end{document}